\documentclass[hidelinks,journal]{IEEEtran}

\usepackage{cite}
\usepackage{latexsym}
\usepackage{verbatim}
\usepackage{amsmath}
\usepackage{amssymb}
\usepackage{theorem}
\usepackage{subcaption}
\usepackage{url}

\usepackage{color,soul}
\sethlcolor{yellow}

\usepackage{xcolor,colortbl}

\usepackage{algorithm}
\usepackage[noend]{algpseudocode}

\newcommand{\algrule}[1][.2pt]{\par\vskip.5\baselineskip\hrule height #1\par\vskip.5\baselineskip}
\usepackage{times}
\usepackage{stfloats}
\usepackage{multirow}
\usepackage{enumerate}
\usepackage{xspace}
\usepackage{longtable}
\usepackage{epstopdf}

\usepackage{graphicx}

\usepackage{threeparttable}  
\newcommand{\specialcell}[2][c]{%
	\begin{tabular}[#1]{@{}c@{}}#2\end{tabular}}

\newtheorem{theorem}{Theorem}

{\theorembodyfont{\rmfamily}
{\bfseries}{\rmfamily}}
{\theorembodyfont{\rmfamily}
\newtheorem{definition}{Definition}{\bfseries}{\rmfamily}}
{\theorembodyfont{\rmfamily}
{\bfseries}{\rmfamily}}
{\bfseries}{\rmfamily}

\newcommand{\cem}{\ensuremath {\texttt{ETA}}{\xspace}}

\newcommand{\params}{\ensuremath {\mathit{I}}{\xspace}}

\newcommand{\EUCMA}{\ensuremath {\mathit{EU}\mhyphen\mathit{CMA}}{\xspace}}

\newcommand{\cemkg}{\ensuremath {\texttt{ETA.Kg}}{\xspace}}
\newcommand{\cemsig}{\ensuremath {\texttt{ETA.Sig}}{\xspace}}
\newcommand{\cemver}{\ensuremath {\texttt{ETA.Ver}}{\xspace}}

\newcommand{\A}{$\mathcal{A}$~}
\newcommand{\F}{$\mathcal{F}$~}

\newcommand{\Rq}{\ensuremath \stackrel{\$}{\leftarrow}\mathbb{Z}_{q}^{*}{\xspace}}

\newcommand{\Ra}{\ensuremath \stackrel{\$}{\leftarrow}{\xspace}}
\newcommand{\Areal}{\ensuremath {\overrightarrow{A}_{\mathit{real}}}{\xspace}}
\newcommand{\Asim}{\ensuremath {\overrightarrow{A}_{\mathit{sim}}}{\xspace}}

\newcommand{\xor}{\oplus}

\newcommand{\as}{\ensuremath {\leftarrow}{\xspace}}

\newcommand{\lh}{\ensuremath {\overrightarrow{\mathcal{L}}}{\xspace}}
\newcommand{\lhr}{\ensuremath {\overrightarrow{\mathcal{L'}}}{\xspace}}
\newcommand{\lsigma}{\ensuremath {\overrightarrow{\mathcal{\sigma}}}{\xspace}}
\newcommand{\lA}{\ensuremath {\overrightarrow{\mathcal{A}}}{\xspace}}

\newcommand{\MM}{\ensuremath{\overline{M}}}
\newcommand{\MT}{\ensuremath{\widetilde{M}}}

\newcommand{\lm}{\ensuremath {\overrightarrow{\mathcal{M}}}{\xspace}}

\newcommand{\hsim}{\ensuremath {\mathit{H}\mhyphen\mathit{Sim}}{\xspace}}

\newcommand{\dl}{\ensuremath {\mathit{DL}}{\xspace}}
\newcommand{\advAdl}{\ensuremath {\mathit{Adv}_{G}^{\dl}(\mathcal{A})}{\xspace}}
\newcommand{\advdl}{\ensuremath {\mathit{Adv}_{G}^{\dl}(t)}{\xspace}}
\newcommand{\advdll}{\ensuremath {\mathit{Adv}_{G}^{\dl}(t')}{\xspace}}

\newcommand{\vv}{\ensuremath {\overrightarrow{v}}{\xspace}}
\newcommand{\ro}{\ensuremath {\mathit{RO}(.)}{\xspace}}

\newcommand{\advcem}{\ensuremath {\mathit{Adv}_{\cem}^{\EUCMA}(t,K',K)}{\xspace}}

\newcommand{\somecs}{\ensuremath {\texttt{SEMECS}{\xspace}}}
	\newcommand{\somecskg}{\ensuremath {\texttt{SEMECS.Kg}}{\xspace}}
	\newcommand{\somecssig}{\ensuremath {\texttt{SEMECS.Sig}}{\xspace}}
	\newcommand{\somecsver}{\ensuremath {\texttt{SEMECS.Ver}}{\xspace}}

\newcommand{\schkg}{\ensuremath {\texttt{Schnorr.Kg}}{\xspace}}
\newcommand{\schsig}{\ensuremath {\texttt{Schnorr.Sig}}{\xspace}}
\newcommand{\schver}{\ensuremath {\texttt{Schnorr.Ver}}{\xspace}}

\newcommand{\advsomecs}{\ensuremath {\mathit{Adv}_{\somecs}^{\EUCMA}(t,K',K)}{\xspace}}

\newcommand{\RNG}{\ensuremath {\texttt{RNG}}{\xspace}}

\mathchardef\mhyphen="2D
\newcommand{\sk}{\ensuremath {\mathit{sk}}{\xspace}}
\newcommand{\pk}{\ensuremath {\mathit{PK}}{\xspace}}

\newcommand{\sgn}{\ensuremath {\texttt{SGN}}{\xspace}}
\newcommand{\sgnkg}{\ensuremath {\texttt{SGN.Kg}}{\xspace}}
\newcommand{\sgnsig}{\ensuremath {\texttt{SGN.Sig}}{\xspace}}
\newcommand{\sgnver}{\ensuremath {\texttt{SGN.Ver}}{\xspace}}

\newcommand{\advAsgn}{\ensuremath {\mathit{Adv}_{\sgn}^{\EUCMA}(\mathcal{A})}{\xspace}}
\newcommand{\advsgn}{\ensuremath {\mathit{Adv}_{\sgn}^{\EUCMA}(t,K',K)}{\xspace}}

\newcommand{\nab}{\ensuremath {\overline{\mathit{E1}}}{\xspace}}
\newcommand{\forge}{\ensuremath {\mathit{E2}}}{\xspace}
\newcommand{\nabb}{\ensuremath {\mathit{\overline{E3}}}{\xspace}}
\newcommand{\suc}{\ensuremath {\mathit{Win}}{\xspace}}

\definecolor{Gray}{gray}{0.85}
\definecolor{LightCyan}{rgb}{0.88,1,1}

\usepackage{hyperref}

\makeatletter
\g@addto@macro{\UrlBreaks}{\UrlOrds}

\newcommand\blfootnote[1]{%
	\begingroup
	\renewcommand\thefootnote{}\footnote{#1}%
	\addtocounter{footnote}{-1}%
	\endgroup
}

\begin{document}

\title{Ultra Lightweight Multiple-time \\ Digital Signature for the Internet of Things Devices}

\author{~Attila~A.~Yavuz,~\IEEEmembership{Member,~IEEE},~Muslum~Ozgur~Ozmen
	\IEEEcompsocitemizethanks{\IEEEcompsocthanksitem \noindent Attila A. Yavuz and Muslum Ozgur Ozmen are with the  Department of Computer Science and Engineering, University of South Florida, Tampa, FL,  USA. \protect\\

		E-mail: attilaayavuz@usf.edu, ozmen@mail.usf.edu
		\IEEEcompsocthanksitem Part of this work is completed when Attila A. Yavuz and Muslum Ozgur Ozmen were with the Department of Electrical Engineering and Computer Science, Oregon State University, Corvallis, OR, USA.}
	\thanks{}
}

\maketitle

\begin{abstract}
Digital signatures are basic cryptographic tools to provide  authentication and integrity in the emerging ubiquitous systems in which resource-constrained devices are expected to operate securely and efficiently. However, existing digital signatures might not be fully practical for such resource-constrained  devices (e.g., medical implants) that have energy limitations. Some other computationally efficient alternatives (e.g., one-time/multiple-time signatures) may introduce high memory and/or communication overhead due to large private key and signature sizes.

In this paper, our contributions are two-fold: First, we develop a new lightweight multiple-time digital signature scheme called  Signer Efficient Multiple-time Elliptic Curve Signature (\somecs), which is suitable for resource-constrained embedded devices. \somecs~achieves optimal signature and private key sizes for an EC-based signature {\em without} requiring any EC operation (e.g., EC scalar multiplication or addition) at the signer. We prove \somecs~is secure (in random oracle model) with a tight security reduction. Second, we fully implemented \somecs~on 8-bit AVR microprocessor with a comprehensive energy consumption analysis and comparison. Our experiments confirm up to 19$\times$ less battery-consumption for \somecs~as compared to its fastest (full-time) counterpart, SchnorrQ, while offering significant performance advantages over its multiple-time counterparts in various fronts. We open-source our implementation for public testing and adoption. \blfootnote{$\copyright$ 2019 IEEE. Personal use of this material is permitted. Permission from IEEE must be obtained for all other uses, in any current or future media, including reprinting/republishing this material for advertising or promotional purposes, creating new collective works, for resale or redistribution to servers or lists, or reuse of any copyrighted component of this work in other works.}

\end{abstract}


\section{Introduction}
\label{sec:Introduction}

Resource-constrained devices (e.g., low-end sensors, smart-cards, RFID-tags) play an important role in emerging ubiquitous systems like Internet of Things and Systems (IoTS) and Wireless Sensor Networks (WSNs). Using service oriented architecture (SOA) further broadens the horizons of IoTS, opening up many applications where resource-constrained devices can participate as service consumers and/or providers.~\cite{TSC_Main,TSC_1,TSC_2} It is vital for such systems to operate securely and efficiently. Hence, it is necessary to provide authentication and integrity for resource-constrained devices. For instance, guaranteeing the integrity and authentication of financial transactions in a smart-card or RFID-tag is critical for any commercial application. However, this is a challenging task due to the memory, processor, bandwidth and battery limitations of these devices.

It is also important to be able to publicly verify the authentication tags produced by resource-constrained devices. This enables any resourceful device (e.g., a laptop or a base station) to publicly audit transactions and system status.

Symmetric cryptography primitives such as Message Authentication Codes (MACs) are computationally efficient and therefore are preferred for resource-constrained devices in small-scale systems. However, such primitives are not scalable for large-distributed systems and are not publicly verifiable~\cite{PKC:WNSInvestigation:Lopez:2006,LogFASFC2012}. They also cannot achieve the non-repudiation property, which is necessary for various applications (e.g., transportation payment systems, medical implants, logical/physical access with tiny devices) that may need public auditing.  

Digital signatures rely on public key infrastructures for the signature verification~\cite{JonathanKatzModernCrytoBook,BroadcastSecureBookPerrig}. They are publicly verifiable, scalable for large systems and achieve non-repudiation. Therefore, they are highly useful authentication tools for security-critical applications such as medical devices, payment systems, secure auditing in embedded devices and security systems (e.g., building access control). However, existing digital signatures also have some limitations that might prevent them to be fully practical for highly resource-constrained devices. 

In the following, we first briefly discuss some prominent digital signature alternatives and their limitations when employed on resource-limited devices. We then present our contributions by summarizing the desirable properties of our scheme, followed by its limitations. 

\subsection{Limitations of Signer Efficient Signatures} \label{subsec:RelatedWork}

The existing digital signature alternatives do not offer small private key size, small signature size, and high efficiency at the signer, at the same time. We elaborate on some of these alternatives below.

\noindent \textbf{Traditional Signatures:} RSA~\cite{RSA}, one of the most well-known signature schemes, is computationally efficient at the verifier's side. However, it requires an {\em expensive operation}\footnote{\small{We refer to operations such as modular exponentiation~\cite{JonathanKatzModernCrytoBook}, elliptic curve scalar multiplication~\cite{ECCGuide} or pairing~\cite{PairingCrypto} as expensive operations.}} at the signer's side and have large key/signature sizes. Hence, it may not be suitable for resource-constrained embedded devices (e.g., medical implants). 

Elliptic Curve (EC) based schemes are highly popular on such devices due to their small key and signature sizes, along with better efficiency compared to RSA~\cite{ECDSA, BLS_asiacrypt, Schnorr91}. Various different curves, and signature schemes on these curves have been proposed, which offer improved computational efficiency and security~\cite{Curve25519Base, FourQ, Kummer, Ed25519, SchnorrQ}. Some of these curves are also implemented in embedded devices~\cite{Ed255198bit, Kummer8bit, FourQ8bit} such as 8-bit AVR microprocessors. However, these signature schemes still require an expensive operation (i.e., EC scalar multiplication) at the signer's side. This requirement may hinder an efficient adoption of these schemes to low-end microprocessors with critical battery limitations (e.g., medical implants).



\noindent \textbf{Online/Offline Signatures:} Many techniques have been proposed to improve the efficiency of traditional signatures. These include online/offline signatures that eliminate expensive operations in signature generation via pre-computed tokens generated offline~\cite{ImprovedDSAEurocrypt94, OnlineOfflineEvenBase1989}. Although these schemes are computationally efficient, they incur large storage overhead to the signer. Later, Shamir et al. in~\cite{OfflineOnline_ImprovedShamir_2001} proposed an improved online/offline signature that is more space efficient. However, by nature, these schemes require linear storage with respect to the number of signatures that can be generated, which is impractical for storage-limited signers. 

\noindent \textbf{One-time Signatures (OTSs):} These schemes rely on one-way functions without trapdoors and offer very efficient computations~\cite{OTS_Lamport_79,HORS_BetterthanBiBa02,BroadAuth:Law:2013:Comparison:AsiaCCS}. Specifically, Lamport~\cite{OTS_Lamport_79} proposed the first one-time signature scheme, where for each bit of the hash of the message, two hash outputs were stored as the public key that resulted in a very large size. Then, in HORS signature scheme~\cite{HORS_BetterthanBiBa02}, special message encoding techniques have been considered to significantly reduce the public while preserving the computational efficiency. Hash-based schemes also offer post-quantum security that is lacked in most of the traditional signatures. On the other hand, they have a large signature and very large public key sizes. Some EC-based OTSs also exist~\cite{Zaverucha} that offer small signature size, but with a trade-off between private key size and signature generation efficiency. Moreover, in OTSs, a private/public key pair can be used only once. This may require costly private key generations at the resource-limited device for each message to be signed. 

\noindent \textbf{Multiple-time Signatures:} For these schemes, after $K$ signatures, the key pair must be re-generated. Therefore, {\em we refer to these schemes as $K$-time signatures}. Inherently, OTSs (e.g., Lamport~\cite{OTS_Lamport_79}, HORS~\cite{HORS_BetterthanBiBa02}) can be used as $K$-time signatures if $K$ key pairs are generated at the key generation phase. Some hash-based multiple-time signatures were proposed~\cite{OTS_r_time_PieprzykWX03, OTS_r_time_HORSED_2006, XMSS, TVHORS} based on HORS signature~\cite{HORS_BetterthanBiBa02}. However, these schemes either suffer from large key/signature sizes~\cite{OTS_r_time_PieprzykWX03, OTS_r_time_HORSED_2006, XMSS}, or provide security only for a short-limited amount of time~\cite{TVHORS}. Some stateless hash-based schemes were also proposed~\cite{SPHINCS}, extending these multiple-time signatures to full-time (traditional) signature schemes. Although it is shown that they can be implemented on resource-constrained devices~\cite{ArmedSPHINCS:Hulsing:PKC:2016}, it is highly computationally costly at the signer's side, and it also requires the transmission of large signatures (e.g., up to 41 KB).

Our proposed scheme also falls into this category and inherits some of the limitations of $K$-time signatures (e.g., after $K$ signatures, the key must be re-generated). However, due to the unique construction of our scheme that leverages Fiat-Shamir transform with compact and efficient elliptic curves, it offers the highest signer efficiency among all the aforementioned $K$-time signatures (including the use of OTSs as $K$-time signatures). For instance, as shown in Table~\ref{tab:AVR}, our scheme outperforms HORS~\cite{HORS_BetterthanBiBa02} (most efficient counterpart) $6\times$ at signature generation and $12 \times$ at signature size on an 8-bit microprocessor.

\noindent \textbf{Lattice-based and Code-based Signatures:} The main advantage of these schemes is {\em post-quantum security}. Although some of these schemes offer computational efficiency, they are still relatively computationally expensive (e.g., require heavy operations such as Gaussian Sampling~\cite{BLISS}) and key/signature sizes might be prohibitive for resource-constrained devices~\cite{BLISS, dilithium, pqsigRM, CodeBasedCourtois, Tachyon}. Since these schemes provide long-term security (security against quantum attackers), they might be ideal for resource-constrained devices in the future, if the sizes are reduced. 
\vspace{1mm}
\noindent\fbox{%
	\parbox{.48\textwidth}{%
		{\em There is a need for a signer efficient digital signature that prioritizes the signer efficiency by achieving optimal private key and signature sizes without requiring any expensive operations for the signature generation.}
	}%
}

\subsection{Our Contribution}
In this paper, we create a new multiple-time signature scheme, which we refer to as {\em \underline{S}igner \underline{E}fficient \underline{M}ultiple-time \underline{E}lliptic \underline{C}urve \underline{S}ignature (\somecs)}. We summarize some desirable properties of \somecs~below. A detailed performance analysis is given in Section~\ref{sec:PerformanceAnalysis}.

$\bullet$~{\em \underline{High Computation \& Energy Efficiency at the Signer}}: \somecs~only requires {\em two hash function calls, a single modular multiplication, and a modular subtraction} to generate a signature. Therefore, its cost is even comparable to symmetric hash-based MACs that are not suitable for large and distributed systems. \somecs~offers very fast signature generation, (1.23 microseconds on an i7 Skylake processor). On a resource-constrained processor, this translates into high energy efficiency. Our experiments confirmed that \somecs~signature generation~has 6$\times$ lower energy consumption compared to its closest counterpart and 118$\times$ lower than Ed25519~\cite{Ed25519, Ed255198bit} on 8-bit AVR microprocessor (see Table~\ref{tab:AVR}).

$\bullet$~{\em \underline{Compact Private Key \& Signature}}: \somecs~only requires storing a 32-byte private key (that can be derived from a 16-byte seed with a PRF) and incurs an additional 32 Bytes to the message as the signature, for $\kappa = 128$-bit security level. \somecs~has two signature components of 32 Bytes, where one of them is used to recover the first 32 Bytes of the message. Therefore, the transmission overhead is just 32 Bytes, that is optimal for EC-based digital signatures (as in BLS~\cite{BLS_asiacrypt}). Thus, \somecs~is also lightweight in terms of signer storage and transmission. Due to its small signature size, \somecs~is also very energy efficient in terms of communication, in addition to its high energy efficiency in signer computation. Moreover, since \somecs~does not require any EC operation at the signer's side, the signer does not need to store any curve parameters and codes. This is specifically important for resource-constrained devices that have limited space for the code (e.g., AVR ATmega 2560 has 256 KB).

\begin{table*}[t!]
	\centering
	\caption{Signer-side performance of \somecs~and its counterparts on 8-bit AVR microprocessor} \label{tab:AVR}
	\vspace{-2mm}
	\begin{threeparttable}
		\begin{tabular}{| c || c | c | c | c |  c | c | }
			\hline
			\textbf{Scheme} & $K$ & \specialcell[]{\textbf{Signature Generation} \\  \textbf{ Time (}CPU cycle\textbf{)}}  &  \specialcell[]{\textbf{Private Key} \\ \textbf{(Byte)}} & \specialcell[]{\textbf{Signature } \\ \textbf{Size (Byte)}} &  \specialcell[]{\textbf{Computation energy} \\ \textbf{(mJ)}} & \specialcell[]{\textbf{Communication energy} \\ \textbf{($\mu$J)}}\\ \hline \hline 
			
			\multicolumn{7}{|c|}{\em \textbf{Full-time signatures}} \\ \hline
			
			SPHINCS~\cite{SPHINCS} & $2^{\kappa}$ & 2 681 600 389 & 1088 & 41000 & 16760.00  & 6115.56 \\ \hline 
			
			
			ECDSA~\cite{ECDSA} & $2^{\kappa}$ &  48 188 992 & 32 & 64 & 301.18 & 9.55 \\ \hline
			
			Ed25519~\cite{Ed25519, Ed255198bit} & $2^{\kappa}$ &  23 211 611  & 32 & 64 & 145.07 & 9.55 \\ \hline 
			
			$\mu$Kummer~\cite{Kummer, Kummer8bit} & $2^{\kappa}$ & 10 404 033 & 32 & 64 & 65.03 & 9.55 \\ \hline 
			
			SchnorrQ~\cite{SchnorrQ, FourQ8bit} & $2^{\kappa}$ & 3 740 000 &  32 & 64 & 23.38 & 9.55 \\ \hline \hline
			
			\multicolumn{7}{|c|}{\em \textbf{$K$-time signatures}} \\ \hline
			
			\multirow{2}{*}{HORS~\cite{HORS_BetterthanBiBa02}} & 1 & \multirow{2}{*}{1 180 618} & \multirow{2}{*}{16} & \multirow{2}{*}{384} & \multirow{2}{*}{7.38} & \multirow{2}{*}{57.28} \\ 
			
			
			& $2^{17}$& & & & &  \\ \hline  
			
			\multirow{2}{*}{HORSE~\cite{OTS_r_time_HORSED_2006}} & 1 & 1 180 618 & 16384 & \multirow{2}{*}{384} & 7.38  & \multirow{2}{*}{57.28} \\ 
			
			
			& $2^{17}$& 19 644 106 & 278528 & & 122.78 &  \\ \hline

			\multirow{2}{*}{XMSS~\cite{XMSS}} & 1 & 10 233 600 & \multirow{2}{*}{16} & 2080 & 63.96 & 310.25 \\ 
			
			
			& $2^{17}$& 101 509 850 &  & 2592 & 634.44 & 386.62 \\ \hline 
			
			\multirow{2}{*}{Zaverucha et al.~\cite{Zaverucha}} & 1 & \multirow{2}{*}{6 250 660} & \multirow{2}{*}{16} & \multirow{2}{*}{48} & \multirow{2}{*}{39.07} & \multirow{2}{*}{7.16} \\ 
			
			
			& $2^{17}$& & & & & \\ \hline \hline
			
			\multirow{2}{*}{\somecs} & 1 & \multirow{2}{*}{ \textbf {195 776} } & \multirow{2}{*}{\textbf {32} } & \multirow{2}{*}{\textbf {32}} & \multirow{2}{*}{ \textbf{1.22}} & \multirow{2}{*}{ \textbf{4.77}} \\ 
			
			
			& $2^{17}$& & & & & \\ \hline 
			
		\end{tabular}
		\begin{tablenotes}[flushleft] \scriptsize{
				\item The cost of hash-based schemes are estimated based on the cost of a single hash operation.
			} 
			
		\end{tablenotes}
	\end{threeparttable}
	\vspace{-5mm}
\end{table*}

$\bullet$~{\em \underline{Open-source Implementation and Comprehensive Analysis}}: We fully implemented \somecs~on a laptop and the signature generation of \somecs~on an 8-bit AVR microprocessor. We open-sourced all of our implementations for broad testing, benchmarking and adoption purposes. We also analyzed and compared the efficiency of \somecs~with a wide variety of efficient signature schemes (see Section~\ref{sec:PerformanceAnalysis}) on both platforms.

$\bullet$~{\em \underline{Provable Security with a Tight Reduction}}: We prove that \somecs~is existentially unforgeable against chosen-message attacks in Random Oracle Model (ROM)~\cite{RandomOracleModel93}. In Section~\ref{sec:SecurityAnalysis}, we show that \somecs~has a tight reduction to the Discrete Logarithm Problem (DLP), without the need for the forking lemma~\cite{ForkingLemma}, as Fiat-Shamir type signatures do. In our security analysis, we exploit the fact that \somecs~is a multiple-time signature, and therefore it has higher security for a limited number of queries (as the nature of multiple-time signatures).

All the above properties show that \somecs~is potentially an ideal alternative to provide authentication and integrity for resource-constrained devices.

\noindent \textbf{Differences of this work with its preliminary version appeared in WiSec 13'~\cite{ETA}}: (i) In this work, we developed a new signature scheme that we refer to as \somecs~that reduces the signature/private key size of our preliminary scheme \cem~\cite{ETA}, and thereby achieves optimal signature and key sizes for an EC-based signature scheme. Moreover, \somecs~generates private key components deterministically, and therefore offers improved security against weak pseudo-random number generators. (ii) In this work, we provided a full-fledged open-source implementation of \somecs~on 8-bit AVR ATmega 2560 microprocessor with a comprehensive energy consumption analysis and comparison. We also gave comprehensive performance comparison of \somecs~with some of the most recent and efficient digital signatures (including but not limited to Ed25519~\cite{Ed25519}, SchnorrQ~\cite{FourQ, SchnorrQ}, SPHINCS~\cite{SPHINCS}, XMSS~\cite{XMSS}, HORS~\cite{HORS_BetterthanBiBa02}). (iii) In this work, we provided an improved security proof that achieves a significantly tighter security reduction compared to that of \cem.

\noindent \textbf{Limitations}: Despite all its merits, \somecs~also has its limitations that are inherent to multiple-time signatures: (i) It can sign up to a pre-determined $K$ messages, but then needs to be bootstrapped. (ii) It is a stateful signature scheme. (iii) In \somecs,  the public key size is linear with respect to $K$. This requires verifiers to be storage resourceful.

\noindent \textbf{Potential Use-cases}: Remark that for our envisioned applications, the signer computational/storage/communication efficiency is much more important than the verifier storage efficiency alone. Furthermore, these applications permit verifiers to be storage resourceful (e.g., a cloud server for medical systems, base stations in WSNs, control centers in cyber-physical systems). Similarly, it is feasible to perform the key generation phase offline in these applications. In the following, we discuss some of the potential applications for \somecs.

Medical implants are equipped with resource-constrained microprocessors (e.g., 8-bit AVR~\cite{IMDProcessor}, as we used in our experiments) that need to report sensitive data to doctors, hospital servers, etc. Symmetric key authentication (MACs) is usually preferred for these systems. However, these mechanisms lack non-repudiation, and public verification that are highly desirable for some medical systems, because of digital forensics and legal issues~\cite{Medical:PersonalHealthDevice:Analysis:Standard:Rubio:2016, MedicalDevice:Survey:2015:Camara2015272}. Thus, there is a need for low-cost public key primitives (e.g., authentication) for these systems~\cite{Ozmen_IOT_SP}. \somecs~can be considered as an ideal alternative for medical implants due to its signer efficiency. Our experiments on 8-bit AVR showed that \somecs~consumes less energy compared to its counterparts. In practice, this translates into a longer battery life that is critical for medical implants.

Additionally, \somecs~is highly suitable to provide authentication for SOA based IoT systems. SOA based IoT infrastructures are comprised of networked, resource-constrained devices~\cite{TSC_Main,TSC_1} that require efficient authentication mechanisms. Some essential applications of SOA based IoT includes but not limited to e-health, smart product management and smart events for emergency management~\cite{TSC_1,TSC_Healthcare_IoT_Services1,TSC_Healthcare_IoT_Services2}. Similarly with medical implants, non-repudiation and public verification are critical for these applications. Moreover, a server (or a broker - i.e., coordinators that operate between the server and the devices) is usually utilized the connected resource-constrained devices to provide these services~\cite{TSC_2}. Servers and brokers are usually equipped with higher-end processors, compared to the IoT devices, that has expandable memories. Therefore, in such SOA based IoT applications, we believe that the server or broker can tolerate the storage of a larger public key in exchange of a significantly higher signer efficiency that translates into longer battery life for resource-constrained devices.

Many secure WSN protocols such as clone detection~\cite{NodeReplication_Journal_ContiPMM11}, secure code dissemination~\cite{SelugeHyun:2008} and secure logging~\cite{SUHaSAFSS11} include a low-end signer that reports to resourceful servers, and base stations. \somecs~can substantially increase the lifespan of WSNs by serving as the authentication mechanism for such protocols. Moreover, \somecs~can be deployed in some token-based logical/physical access control systems.

\section{Preliminaries}
\label{sec:Prelim}
In this section, we first give our notation and definitions. We then describe our system and security models. 

\subsection{Definitions and Algorithms} \label{subsec:DefSecModel}
\noindent\textbf{Notation.} $||$ denotes the concatenation operation.
$|r|$ denotes the bit length of variable $r$. $r \stackrel{\$}{\leftarrow}\mathcal{S}$ denotes that variable $r$ is
randomly and uniformly selected from set $\mathcal{S}$.  We denote by
$\{0,1\}^{*}$ the set of binary strings of any finite length. $\mathcal{A}^{\mathcal{O}_0,\ldots,\mathcal{O}_{i}}(\cdot)$
denotes algorithm $\mathcal{A}$ is provided with oracles
$\mathcal{O}_0,\ldots,\mathcal{O}_{i}$. For example,
$\mathcal{A}^{\mathit{SGN.Sig}_{\sk}}(\cdot)$ denotes that algorithm
$\mathcal{A}$ is provided with a {\em signing oracle} of signature
scheme $\mathit{SGN}$ under a private key \sk. $H_i: \{0,1\}^{*} \rightarrow \mathbb{Z}_{q}^{*}, i\in\{0,1\}$ are distinct Full Domain Hash Functions~\cite{FullDomainHashExact2000}, where $q$ is a large prime.

\begin{definition}
	\label{Def:CEMGenericDefiniton}
	A $K$-time signature scheme \sgn~is comprised of a tuple of three algorithms
	$(\mathit{Kg},$ $\mathit{Sig},\mathit{Ver})$ defined as follows:

	\begin{enumerate}[-]
		\item $(\sk,\pk)\leftarrow
		\sgnkg(1^{\kappa},K)$,: The key generation
		algorithm takes the security parameter $1^{\kappa}$ and the maximum
		number of messages to be signed $K$ as the input. It returns a private/public
		key pair $(\sk_0,\pk)$ as the output.
		
		\item $\sigma_{j}\leftarrow	\sgnsig(\sk_j,M_j)$: The signature generation algorithm takes the private key $\sk_j,~0\leq j \leq K-1$ and a message $M_j$ to be signed as the input. It returns a signature
		$\sigma_{j}$ on $M_j$ as the output, and then updates $\sk_j$ to $\sk_{j+1}$. 
		
		\item $b\leftarrow \sgnver(\pk,M_j,\sigma_{j})$: The signature verification algorithm takes $\pk$, message $M_j$ and its corresponding signature $\sigma_{j},~0\leq j \leq K-1$ as the
		input.  It returns a bit $b$, with $b=1$ meaning {\em valid}, and
		$b=0$ otherwise.
	\end{enumerate}
\end{definition}

\somecs, and its preliminary version \cem~in~\cite{ETA}, are inspired from the Schnorr signature scheme~\cite{Schnorr91}, which is described in the algorithm below.

\vspace{-2mm}
\begin{algorithm}[h!]
	\caption{Schnorr Signature Scheme}\label{alg:schnorr}
	\hspace{5pt}
	\begin{algorithmic}[1]
		
		\Statex   $\underline{(y,Y) \leftarrow \schkg(1^{\kappa})}$: 
		\vspace{3pt}
		\State Generate large primes $q$ and $p>q$ such that $q | (p-1)$.
		\State Select a generator $\alpha$ of the subgroup $G$ of order $q$ in $\mathbb{Z}_{p}^{*}$.		
		\State \Return a private/public key pair $(y\stackrel{\$}{\leftarrow}\mathbb{Z}_{q}^{*},Y\leftarrow
		\alpha^{y} \bmod p)$ and system-wide parameter $\params\as(q,p,\alpha)$ as the output.
	\end{algorithmic}
	\algrule
	
	\begin{algorithmic}[1]
		\Statex $\underline{(s,e)\leftarrow \schsig(y,M)}$: Given $y$, compute signature $\sigma$ on a message $M$ as follows:
		\vspace{3pt}
		\State $R\as\alpha^{r} \bmod p$.
		\State $e\as H_0(M||R)$.
		\State $s\as(r-e\cdot y) \bmod q$, where $r\stackrel{\$}\as \mathbb{Z}_{q}^{*}$.
		\State \Return $\sigma = (s,e)$.

	\end{algorithmic}
	\algrule
	
	\begin{algorithmic}[1]
		\Statex $\underline{b\leftarrow \schver(Y,M,\langle	s,e \rangle)}$: 
		\vspace{3pt}
		\State $R'\as Y^{e}\alpha^{s} \bmod p$.
		\If{$e=H_0(M||R')$} \Return $b=1$
		\Else~\Return $b=0$
		\EndIf
	\end{algorithmic}
\end{algorithm}

\subsection{Models} \label{subsec:SecModel}
We give our system and security models as below.

\noindent \textbf{System Model}: There are two types of entities in the system. 

\begin{enumerate}
	
	\item {\em Resource-constrained Signers}: Signers are storage, computational, bandwidth and power limited devices (e.g., medical implants, wireless sensors, RFID-tags). The objective of \somecs~is to minimize the cryptographic overhead of signers.
	
	\item {\em Resourceful Verifiers}: Storage resourceful verifiers (e.g., a laptop, base station) that can be any (untrusted) entity.
\end{enumerate}

	
We assume that the key generation/distribution is performed {\em offline} before deployment. For instance, a key generation center can generate private/public keys and distribute them to each entity in the system. Otherwise, the signer can also perform the key generation, before deployment, when it does not have any battery limitations.



\noindent \textbf{Security Model}: A standard security notion for a signature scheme is Existential Unforgeability under Chosen Message Attack (\EUCMA)~\cite{ModernCryptoBellareRogaway}. We define $K$-time {\em EU-CMA} experiment (in random oracle model~\cite{RandomOracleModel93}) for \sgn~as below. In this experiment, Adversary $\mathcal{A}$ is provided with two oracles: (i) A {\em random oracle} \ro~from which \A can request the hash of any message $M$ of their choice up to (polynomially unbounded) $K'$ messages. (ii) A signing oracle $\sgnsig_{sk}(.)$~from which \A can request a \sgn~signature on any message $M$ of their choice up to (pre-determined constant) $K$ messages.
%
%
\begin{definition}
	\label{Def:CEMSecurityModel}
	{\em EU-CMA experiment} for \sgn~is as follows:
	
	\noindent Experiment $\mathit{Expt}_{\sgn}^{\EUCMA}(\mathcal{A})$
	\begin{enumerate}[]
		\setlength{\parskip}{0pt}
		\setlength{\parsep}{0pt}
		
		\item $(sk_0,PK)\leftarrow
		\sgnkg(1^{\kappa},K)$,
		
		\item $(M^{*},\sigma^{*})\leftarrow \mathcal{A}^{\ro,\sgnsig_{sk}(.)}(PK)$,
		
		\item If $\sgnver(PK,M^{*},\sigma^{*})=1$ and $M^{*}$ was not queried to \sgnsig, return $1$, else, return $0$.
	\end{enumerate}
	\noindent
	
	The EU-CMA-advantage of \A is defined as
	\begin{eqnarray*}
		\advAsgn=Pr[\mathit{Expt}_{\sgn}^{\EUCMA}(\mathcal{A})=1].
	\end{eqnarray*}

	The EU-CMA-advantage of \sgn~is defined as
	\begin{eqnarray*}
		\advsgn=\max_{\mathcal{A}}\{Adv_{\sgn}^{\EUCMA}(\mathcal{A})\},
	\end{eqnarray*}
	where the maximum is over all $\mathcal{A}$ having time complexity
	$t$, making at most $K'$ queries to \ro~and at most $K$ queries to \sgn.
\end{definition}



The security of \somecs~relies on the intractability of {\em Discrete Logarithm
	Problem (DLP)}~\cite{ModernCryptoBellareRogaway}, which is defined below.

\begin{definition} \label{Def:DL}
	Given a cyclic group $G$ of order prime $q$ and a generator $\alpha$ of
	$G$, let $\mathcal{A}$ be an algorithm that returns an element of
	$\mathbb{Z}_{q}^{*}$. Consider the following experiment: 
	
	\vspace{1mm}
	
	\noindent Experiment $\mathit{Expt}_{G}^{\dl}(\mathcal{A})$
	\begin{enumerate}[]
		\setlength{\parsep}{0pt}
		\item $y \stackrel{\$}{\leftarrow}\mathbb{Z}_{q}^{*}$,
		
		\item $Y\leftarrow \alpha^{y} \bmod p$,
		
		\item $y'\leftarrow \mathcal{A}(Y)$,
		
		\item If $\alpha^{y'} \bmod p = Y$, return $1$, else, return $0$.
	\end{enumerate}
	The {\em DL-advantage of $\mathcal{A}$} in this experiment is defined as
	\begin{eqnarray*}
		\advAdl=Pr[\mathit{Expt}_{G}^{\dl}(\mathcal{A})=1].
	\end{eqnarray*}
	The {\em DL-advantage of $(G,\alpha)$} in this experiment is defined as
	\begin{eqnarray*}
		\advdl=\max_{\mathcal{A}}\{\advAdl\},
	\end{eqnarray*} where the maximum is over all $\mathcal{A}$ having
	time complexity $t$.
\end{definition}


\section{The Proposed Scheme}
\label{sec:ProposedSchemes}

Some DLP-based signatures (e.g., ECDSA~\cite{ImprovedDSAEurocrypt94}, Schnorr~\cite{Schnorr91}) can eliminate expensive operations from the signature generation by pre-computing the component $R=\alpha^{r} \bmod p$ for a random $r\Rq$ during the key generation. The signer stores $(r,R)$ and then use them to compute signatures during the online phase, without any expensive operation. However, this approach incurs linear storage to the signer's side (i.e., one token per message). 

It is highly desirable to construct a multiple-time signature scheme, which has constant signer storage and yet avoids expensive operations. However, this is a challenging task due to the nature of the aforementioned schemes. That is, in these schemes, the token $R$ is directly used during the signature computation and therefore its storage cannot be trivially off-loaded to the verifier's side. This forces the signer either to store or to compute a token for each message.

\subsection{Preliminary Scheme: Efficient and Tiny Authentication}
In our preliminary work  {\em Efficient and Tiny Authentication (ETA)}~\cite{ETA}, we designed a signature scheme that can shift the storage of ephemeral public keys to the verifier's side without disrupting the security and verifiability of signatures. We outline our preliminary scheme  \cem~in Algorithm~\ref{alg:eta} for the sake of completeness.

In the following, we focus on our newly proposed \somecs~digital signature scheme and also highlight the differences between \cem~and our improved scheme \somecs. 

\subsection{Signer Efficient Multiple-time Elliptic Curve Signature (\somecs)} \label{subsecOMECS}
We first discuss the challenges of eliminating ephemeral keys  from the signature generation in Schnorr-like signatures, which is an important step to achieve signer optimal elliptic curve signatures. We then explain our strategies in \somecs~towards addressing these challenges.

\begin{algorithm}[t!]
	\caption{Efficient and Tiny Authentication (\cem) Scheme}\label{alg:eta}
	\hspace{5pt}
	\begin{algorithmic}[1]
		
		\Statex   $\underline{(\sk_0,\pk,\params)\as\cemkg(1^{\kappa},K)}$: 
		\vspace{3pt}
		\State $(y,Y,\langle q,p,\alpha \rangle) \as \schkg(1^{\kappa})$.
		\State $r_0\Rq$
		\For{$j=0,\ldots,K-1$}
		\State $R_j\as \alpha^{r_j} \bmod p$.
		\State $r_{j+1}\as H(r_{j})$.
		\State Generate verification tokens as $v_j\as H(R_j)$.
		\EndFor
		\State \Return The private and public key, as $\sk_0\as(y,r_0)$ and $\pk\as(Y,\vv=v_0,\ldots,v_{K-1})$, respectively.
	\end{algorithmic}
	\algrule
	
	\begin{algorithmic}[1]
		\Statex $\underline{\sigma_j\as \cemsig(\sk_{j},M_j)}$: Given $\sk_j=(y,r_j)$, compute signature $\sigma_j$ on a message $M_j$ as follows:
		\vspace{3pt}
		\State $x_j\Ra\{0,1\}^{\kappa}$.
		\State $e_j\as H(M_j||j||x_j)$.
		\State $s_j\as r_j-e_j\cdot y \bmod q$.
		\State The signature $\sigma_j$ on $M_j$ is $\sigma_j\as (s_j,x_j,j)$.
		\State Update $r_j$ as $r_{j+1}\as H(r_j)$, erase $r_j$ (to save memory).
		\If{$j> K-1$} \Return $\perp$ (i.e., the limit on the number of signatures is exceed).
		\Else~\Return $\sigma_j$
		\EndIf
	\end{algorithmic}
	\algrule
	
	\begin{algorithmic}[1]
		\Statex $\underline{b \as \cemver(\pk,M_j,\sigma_j)}$: If $j \ge K$ then return $b=0$ and {\em abort}. Otherwise, continue as following:
		\vspace{3pt}
		\State $R_j'\as Y^{H(M_j||j||x_j)}\cdot \alpha^{s_j}$.
		\If{$v_j=H(R_j')$} \Return $b=1$
		\Else~\Return $b=0$
		\EndIf
	\end{algorithmic}
\end{algorithm}

\subsubsection{Challenges of Removing Ephemeral Key from Signature Generation}
In Schnorr-like signatures~\cite{ECDSA, Ed25519, SchnorrQ}, an expensive operation is required to compute the ephemeral key $(R=\alpha^{r} \bmod p,r\Rq)$. This ephemeral key is an essential part of the signature generation and proof of security, and therefore it is a challenging task to remove it from signing without disrupting the security. For example, $R$ is committed to the signature as $s\as r-H(M||R)\cdot y \bmod q$ in Schnorr signatures~\cite{Schnorr91}. The ephemeral key enables programming of random oracle and also used in Forking Lemma~\cite{ForkingLemma} in the security proof of Schnorr-like signatures~\cite{Ed25519, SchnorrQ}.

\subsubsection{Eliminating Expensive Operations from Signature Generation} We first pre-compute $K$ ephemeral keys as $r_j\as H_0(y||j)$,  $R_j\as \alpha^{r_j} \bmod p$  and store their hash commitments at the verifier as  $\beta_j \as H_1(R_j)$  for $j=0,\ldots,K-1$ (Steps 3-7 in Algorithm~\ref{alg:somecs} $\somecskg$). This permits the derivation of $r_i$ to be used in signature $s_j$ deterministically without requiring any expensive operation, which will later to be verified by its corresponding $\beta_j$. Since $R_j$ is not required in the signature generation, we avoid  expensive operations, but only rely on a few hash calls and a single modular addition/multiplication. 

Our next step is to ensure that the correctness and provable security are still achieved in the absence of the ephemeral key in the signature generation. In \cem~\cite{ETA}, we mimicked the role of $R_j$ in $e_j$ by replacing it with a random number $x_j\as\{0,1\}^{\kappa}$ as $e_j\as H(M_j||j||x_j)$ (Steps 1-2 in Algorithm~\ref{alg:eta} $\cemsig$). However, this requires the explicit transmission of an extra $\kappa$-bit randomness and therefore is not optimal in terms of signature size. Moreover,  this random number must be generated online, so requires a strong random number  generator to be present in a low-end device. 

In the following (Section \ref{subsubsec:Compactkey}), we first outline how \somecs~improves the signature generation of \cem~by reducing the private key and signature sizes.  We then elaborate on how \somecs~achieves the correctness and a tight security reduction in Section \ref{subsubsec:VerificationReduction}. 

\begin{algorithm}[t!]
	\caption{Signer Efficient Multiple-time Elliptic Curve Signature (\somecs) Scheme}\label{alg:somecs}
	\hspace{5pt}
	\begin{algorithmic}[1]
		
		\Statex   $\underline{(\sk_0,\pk)\as \somecskg(1^{\kappa},K)}$: 
		\vspace{3pt}
		\State $(y,Y,\langle q,p,\alpha \rangle) \as \schkg(1^{\kappa})$. 
		\For{$j=0,\ldots,K-1$}
		\State $r_j\as H_0(y||j)$.
		\State $R_j\as \alpha^{r_j} \bmod p$.
		\State $z_j\as H_1(y||j)$.
		\State $\gamma_j\as z_j \xor H_0(R_j)$.
		\State $\beta_j \as H_1(R_j)$.
		\EndFor
		\State \Return $\sk_0\as y$ and $\pk\as(Y,\alpha,\vv=( \langle \gamma_0,\beta_0 \rangle,\ldots, $ $\langle \gamma_{K-1},\beta_{K-1} \rangle,K)$.
	\end{algorithmic}
	\algrule
	
	\begin{algorithmic}[1]
		\Statex $\underline{\sigma\as \somecssig(\sk_j, M_j)}$: Given $\sk_j=(y,j)$ compute the signature as follows: 
		\vspace{3pt}
		\If{$|M_j| < |q|$ } set $(\MM_j = M_j, \MT_j=0)$, 
		\Else~ split $M_j$ into two as $(\MM_j||\MT_j)$ such that $|\MM_j|=|q|$.
		\EndIf
		\State $r_j\as H_0(y||j)$.
		\State $z_j\as H_1(y||j)$.
		\State $c_j\as\MM_j \xor z_j$.
		\State $e_j \as H_0(c_j||\MT_j)$.
		\State $s_j\as r_j-e_j\cdot y \bmod q$.
		\If{$j> K-1$} \Return $\perp$ (i.e., the limit on the number of signatures is exceeded).
		\Else~\Return The signature $\sigma_j$ on $M_j$ is $\sigma_j\as (s_j,c_j)$, where the sender transmits $(\sigma_j,\MT_j)$ to the receivers.
		\EndIf
	\end{algorithmic}
	\algrule
	
	\begin{algorithmic}[1]
		\Statex $\underline{b \as \somecsver(\pk,\MT_j,\sigma_j)}$: If $|c_j| > |q|$ or $j \ge K$ then \somecsver~return 0 and {\em aborts}. Otherwise, continue as following:
		\vspace{3pt}
		\State $ R_j'\as Y^{H_0(c_j||\MT_j)}\cdot \alpha^{s_j}  \bmod p$.
		\If{$\beta_j \neq H_1(R_j')$}  \Return $b=0$.
		\Else~\Return $b=1$ and recover the message $M_j$ as follows:
		\State $ \MM_j \as \gamma_j \xor H_0(R_j') \xor c_j$.
		\If{$\MT_j=0$} set $M_j=\MM_j$.
		\Else~set $M_j=(\MM_j||\MT_j)$.
		\EndIf
		\EndIf  
	\end{algorithmic}
\end{algorithm}

\subsubsection{Achieving Compact Key and  Signature Sizes} \label{subsubsec:Compactkey}
Our idea is to  embed randomness into the message itself by creating a ``randomized message recovery" strategy, thereby avoiding an explicit transmission of randomness. 

We first split message $M_j$ into two pieces as  $(\MM_j||\MT_j)$ such that $|\MM_j|=|q|$ and $\MT_j$ is the rest of message. If $|M_j| < |q|$ then we simply set $\MM_j = M_j$ and $\MT_j=0$ (Steps 1-2 in Algorithm~\ref{alg:somecs} $\somecssig$). We then deterministically derive $z_j\as H_1(y||j)$, generate a randomness as  $c_j\as\MM_j \xor z_j$ and compute the hash of the message as  $e_j \as H_0(c_j||\MT_j)$. Finally, we compute  $s_j\as r_j-e_j\cdot y \bmod q$ (Step 7 in Algorithm~\ref{alg:somecs} $\somecssig$).

Our signature $\sigma_j$ on $M_j$ is $\sigma_j = (s_j,c_j)$, where the sender transmits $(\sigma_j,\MT_j)$ to the receivers. Remark that, we only transmit  $c_j$ that carries $|q|$-bit part of the message since  $c_j\as\MM_j \xor z_j$. Therefore, the only component of the signature that introduces cryptographic transmission overhead is $s_j \in \mathbb{Z}_{q}^{*}$, which is optimal for an elliptic curve based signature scheme\footnote{$c_j$ does not offer confidentiality. After $\sigma_j=(s_j,c_j)$ is released, the message and $z_j$ are publicly recovered to permit signature verification.}. This is as small as some of the most compact signatures (e.g., BLS~\cite{BLS_asiacrypt}) but without requiring expensive operations at the signer's side. Morever, it is also smaller than SchnorrQ~\cite{FourQ, SchnorrQ} and \cem~\cite{ETA} that transmit $e_j$ and $x_j$, respectively, as an extra information on top of $s_j$. 

\somecs~achieves a small private key $y \in \mathbb{Z}_{q}^{*}$, which is identical to that of traditional Schnorr-like signatures~\cite{ECDSA, Ed25519, SchnorrQ} and only a half of the size that of \cem's private key. The small and constant private key size is achieved by generating the random values with a deterministic function (e.g., a hash function) just using a seed value ($y$). Therefore, the signer doesn't need to store all the random values generated at key generation, but only stores the seed and deterministically derives all random values from it ($\somecssig$ Step 3-4). Moreover, unlike \cem, \somecs~signature generation does not require any fresh randomness and therefore avoids potential hurdles of weak pseudo-random number generators on the signer device~\cite{ATTACKprng, PRNG-Nguyen}. 

\begin{figure*}[!t]
	\centering
	\includegraphics[width=.88\linewidth]{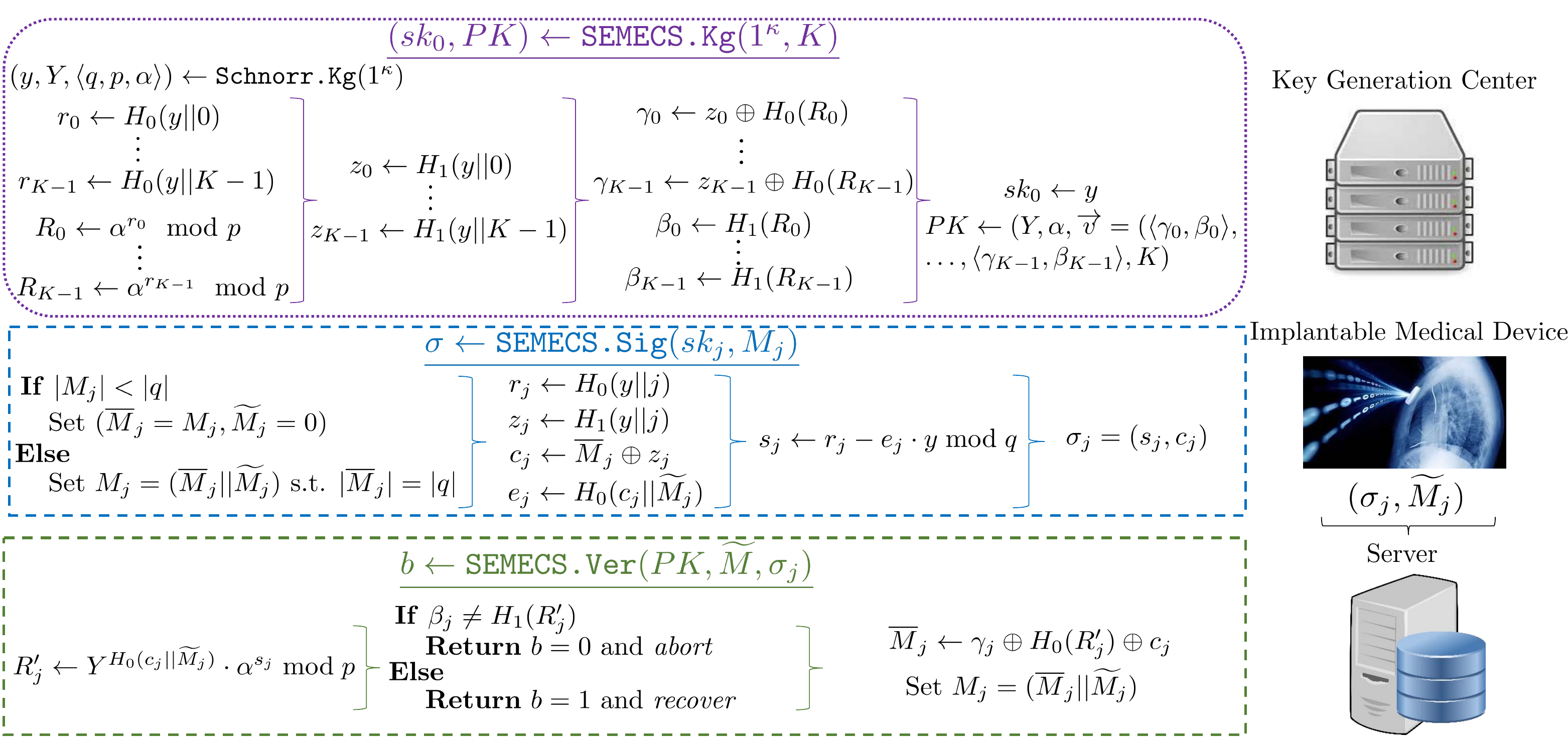}
	\caption{High-level description of \somecs~algorithms.}
	\label{fig:somecs}
	\vspace{-5mm}
\end{figure*}

\subsubsection{Signature Verification and Tight Security Reduction}\label{subsubsec:VerificationReduction}

The verifier first checks the range of randomness $c_j \in \mathbb{Z}_{q}^{*}$ and the limit on number of permitted signatures. The verifier then computes $R_j'\as Y^{H(c_j||\MT_j)}\cdot \alpha^{s_j}$ and checks whether it matches with $\beta_j=H_1(R_j') \in \pk$. If it does not hold, the verifier returns $b=0$. Otherwise, the verifier returns $b=1$ and uses auxiliary value $\gamma_j$ to recover the $q$-bit piece of message $\MM_j$ from $c_j$ as $ \MM_j \as \gamma_j \xor H_0(R_j') \xor c_j$ forming the original  message as $M_j = (\MM_j||\MT_j)$.

Note that the verifier should either know which public key component ($\beta_j$) it should use at $\somecsver$ Step 2 or have a simple search operation among all $\beta$s to see if there is one that matches the calculated $H_1(R_j')$. Therefore, there is a trade-off between a verifier computation and transmission overhead. However, both of these costs are almost negligible. Since $j$ is a value up to $K$ ($K = 2^{17}$ in our experiments), the transmission of it only incurs 2-3 Bytes of extra overhead. If the verifier computation is preferred, this only adds an overhead of a binary search operation, that has a complexity of $log_2(K)$. In the binary search option, we basically assume that the verifier stores the public key sorted, and after the value $H_1(R_j')$ is calculated, binary search is made on sorted $\beta$s.

We now elaborate the design rationale behind the use of two separate verification tokens $(\beta_j,\gamma_j)$ in \somecs, as opposed to only one token $v_j$ in \cem, for $j=0,\ldots,K-1$. 

(i) In Schnorr-like schemes, the randomness incorporated into message hashing is released with $s_j$ but {\em not before}. This is useful to construct an indistinguishable simulation in the security proof of Schnorr-like signatures\footnote{ In our security proof for \somecs~in Theorem \ref{the:Theorem1}, the simulator \F programs random oracle \ro~such that the probability that adversary \A querying \ro~on $c_j||M_j$ {\em before} querying it to the signature oracle $\somecssig_{sk}$ is as difficult as random guessing $c_j \in \mathbb{Z}_{q}^{*}$. Hence, the probability that simulator \F aborts during the query phase is negligible in terms of $\kappa$ (see success probability analysis in Theorem \ref{the:Theorem1}). }. In \somecs, $c_j$ that randomizes the message hash as $H_0(c_j||\MT_j)$, is computed from $z_j$ as  $c_j\as\MM_j \xor z_j$. Our idea is to store $z_j$ at the verifier's side as $\gamma_j \as z_j \xor H_0(R_j)$ so that it can  be recovered {\em only after} $s_j$ is released. We avoid an online transmission of $z_j$ but yet randomize the message hash via $c_j$ including $q$-bit part of the message $\MM_j$ (with no extra transmission overhead). After $\sigma_j = (c_j,s_j)$ is released, the verifier computes  $z_j$ from $\gamma_j$ via $H_0(R_j)$. Note that $\beta_j=H_1(R_j)$ does not reveal $z_j$ but yet permits the verification of $ R_j'\as Y^{H_0(c_j||\MT_j)}\cdot \alpha^{s_j}  \bmod p$.

(ii) In \somecs, we present an improved security proof with a reduction to DLP with a much tighter bound compared to that of \cem. The security of \cem~is reduced to Schnorr signatures, whose security proof relies on Forking Lemma~\cite{ForkingLemma}. Intuitively, if there is an adversary \A making at most $K'$ \ro~queries, and forging signatures with probability $\epsilon$, then the Forking Lemma states that one can compute discrete logarithms with constant probability by rewinding the forger $O(K'/\epsilon)$ times. Therefore, the security reduction loses a factor $O(K')$ that can be very large~\cite{Seurin2012}. 

Our key observation is that, since \somecs~is a $K$-time signature with pre-determined ephemeral public keys, we can avoid using Forking Lemma and obtain a reduction to DLP. That is, the hash of ephemeral keys are committed at the key generation phase as $\{\beta_j\}_{j=0}^{K-1} \in \pk$. At the forgery phase, if \A outputs a forgery on \pk~as $(M^{*},$ $\sigma^{*})$, where $\sigma^{*}=(s_{j}^{*},c_{j}^{*}),~0\leq j \leq K-1$, by validity condition, this forgery has to be on  a $\beta_j \in \pk$. Therefore, \F can extract private key $y$ {\em without} a need of rewinding \A. This permits us to avoid a large factor of $O(K')$ but only need a small constant factor $O(K)$ in our security reduction. We stress that this is possible due to special $K$-time nature of \somecs, but it does {\em not} apply to polynomially unbounded Schnorr signature variants as proven in~\cite{Seurin2012}.

The detailed description of \somecs~is given at Algorithm~\ref{alg:somecs} and further outlined in Figure~\ref{fig:somecs}.


\section{Security Analysis}
\label{sec:SecurityAnalysis}
We prove that \somecs~is a $K$-time {\em EU-CMA} signature scheme in Theorem ~\ref{the:Theorem1} (in the random oracle model~\cite{RandomOracleModel93}). We ignore terms that are negligible in terms of $\kappa$.

\begin{theorem} \label{the:Theorem1}

$\advsomecs \le  \advdll$, where  $t'=O(t)+(2K)\cdot O(\kappa^{3})+(6K+K')\cdot\RNG$.


\end{theorem}
\noindent {\em Proof:} Let \A be a \somecs~attacker. We construct a {\em DL-attacker} \F that uses $\mathcal{A}$ as a sub-routine. That is, we set $(y'\Rq, Y'\as \alpha^{y'} \bmod p)$ as defined in $\mathit{DL\mhyphen experiment}$ (i.e., Definition~\ref{Def:DL}) and then run the simulator \F  by Definition~\ref{Def:CEMSecurityModel} (i.e., \EUCMA~experiment) as follows:

\vspace{2mm} \noindent \underline{{\em Algorithm $\mathcal{F}(Y')$}}
\begin{enumerate}[-]
\setlength{\itemsep}{2pt}
  \setlength{\parskip}{1pt}
  \setlength{\parsep}{1pt}
\item \underline{{\em Setup:}}  \F keeps three lists \lm,~\lh,~and \lhr, all initially empty. \lm~is a message list that records each $M_j$ queried to \cemsig~oracle. $\lh[j]$ and $\lhr[j]$~record $(M_j,i)$ queried to \ro~oracle and its corresponding \ro~answer $(h_j,i)$, respectively, for cryptographic hash functions $H_i, i \in \{0,1\}$.  $(h_j,i)\as \lhr[M_j,i]$ denotes the retrieval of \ro~oracle answer of $(M_j,i)$ that has been queried before. If $(M_j,i)$ has not been queried before then $\perp \as \lhr[M_j,i]$. \F sets counters $(l\as0,n\as0)$ and continues as follows: 

     \begin{enumerate}[$\bullet$]

     \item   {\em $h\as\hsim(M,l,\lh,\lhr,i)$}: \F implements a function \hsim~to handle \ro~queries. That is, cryptographic functions $H_i, i \in \{0,1\}$ are modeled as random oracles via \hsim. If $\exists{j}:(M,i)=\lh[j]$ then \hsim~returns $\lhr[j]$. Otherwise, it~returns $h\Rq$ as the answer for given $H_i$, assigns $(\lh[l]\as (M,i),~\lhr[l]\as (h,i))$ and $l \as l+1$.

     \item \F creates a simulated \somecs~public key \pk~as follows:
      \begin{enumerate}[-]
      \item $Y\as Y'$,
      \item For $j=0,\ldots,K-1$,
      \begin{enumerate} [a)]
      	\item $(s_j,e_j,z_j)\Ra \mathbb{Z}_{q}^{*}$ 
      	\item  $R_j\as Y^{e_j}\cdot \alpha^{s_j} \bmod p$ 
      	 \item $\gamma_j\as z_j \xor \hsim(R_j,l,\lh,\lhr,0)$ 
      	 \item $\beta_j \as\hsim(R_j,l,\lh,\lhr,1) $
      	\end{enumerate}
     
      \item  $\pk\as(Y,\alpha,\vv=( \langle \gamma_0,\beta_0 \rangle,\ldots, $ $\langle \gamma_{K-1},\beta_{K-1} \rangle,K)$.
      \end{enumerate}

\end{enumerate}

\item \underline{\em Execute $(M^{*},\sigma^{*})\as\mathcal{A}^{\ro,\somecssig_{sk}(.)}(\pk)$}: 

  \begin{enumerate}[$\bullet$]

    \item    \underline{Queries:} \F handles \A's  queries as follows:

         {\em (i)} \A queries \ro~on a message $M$ for $H_i, i\in\{0,1\}$.  \F returns $h\as\hsim(M,l,\lh,\lhr,i)$.

        {\em (ii)} \A queries \somecssig~oracle on a message $M_n$. If $n>K-1$ then \F rejects the query (i.e., the query limit is exceeded). Otherwise, \F  continues as follows:
        
         \begin{enumerate}[a)]
        	\setlength{\itemsep}{3pt}
        	\setlength{\parskip}{1pt}
        	\setlength{\parsep}{1pt}
        	
        	\item If $|M_n| < |q|$ set $(\MM_n = M_n, \MT_n=0)$, else split $M_n$ into two  as $(\MM_n||\MT_n)$ such that $|\MM_n|=|q|$.
        	
        	\item   \F generates $c_n \as z_n \xor \MM_n$ and checks if $(c_n||M_n,0)\in \lh$. If it holds then \F {\em aborts} (i.e., the simulation fails). Otherwise, \F continues as follows. 
        	
        	\item  \F inserts $(\lh[l]\as (c_n||M_n,0),\lhr[l]\as (e_n,0)$). 
        	
        	\item  \F returns  $\sigma_n \as (s_n,c_n)$ to \A, sets $\lm[n]\as M_n$ and then increments $(n\as n+1,l\as l+1)$.
        	\end{enumerate}

    \item    \underline{Forgery of \A:} Eventually, \A outputs a forgery on \pk~as $(M^{*},$ $\sigma^{*})$, where $\sigma^{*}=(s_{j}^{*},c_{j}^{*}),~0\leq j \leq K-1$. By definition \ref{Def:CEMSecurityModel}, \A wins the $K$-time \EUCMA~experiment for \somecs~if $\somecsver(\pk,$ $M^{*},\sigma^{*})=1$ and  $M^{*} \notin \lm$ hold. If these conditions hold, \A returns $1$, else, returns $0$.
  \end{enumerate}

\item \underline{Forgery of \F}: If \A loses in the $K$-time \EUCMA~experiment for \somecs, \F also loses in the \dl~experiment, and therefore \F {\em aborts} and returns $0$. Otherwise, if $(c_{j}^{*}||M^{*}) \in \lh$ then \F {\em aborts} and returns $0$ (i.e., \A wins the experiment without querying \ro~oracle).  Otherwise, \F sets $s_{j}^{*} \as \lhr[c_{j}^{*}||M^{*}]$ and continues as follows:\\

Recall that $R_{j} \equiv Y^{e_{j}} \cdot \alpha^{s_j} \bmod p$ holds, where $0\leq j \leq K-1$. Moreover, since $\somecsver(\pk,M^{*},\sigma^{*})=1$ holds, $R_{j} \equiv Y^{{e_j}^{*}}\cdot \alpha^{s_{j}^{*}} \bmod p$ also holds. Therefore, we write the following equations:
\begin{eqnarray*}
R_{j} \equiv Y^{e_j}\cdot \alpha^{s_j} \bmod p, \\
R_{j} \equiv Y^{{e_j}^{*}}\cdot \alpha^{s_{j}^{*}} \bmod p,
\end{eqnarray*}

\F then extracts $y'=y$ by solving the below modular linear equations (note that only unknowns
are $y$ and $r_j$), where $Y=Y'$ as defined in simulation:
\begin{eqnarray*}
 r_j      & \equiv & y'\cdot e_j+ s_j \bmod q,\\
 r_j  & \equiv & y'\cdot e_{j}^{*} + s_{j}^{*} \bmod q,
\end{eqnarray*}

Note that $Y'\equiv \alpha^{y'} \bmod p$ holds, since \A's forgery is valid and non-trivial on $Y'=Y$. Therefore, by Definition \ref{Def:DL}, $\mathcal{F}$ wins the $\mathit{DL\mhyphen experiment}$.
    \end{enumerate}

The execution time and probability analysis of the above experiment are as follows:

\noindent \underline{{\em Execution Time Analysis}}: In this experiment, the running time of \F is that of \A plus the time it takes to respond $q_H$ \ro~queries and $K$ \cemsig. 
\begin{itemize}
 \item {\em Setup phase:} \F draws $3K$ random numbers, performs $2K$ modular exponentiations, $K$ XOR operations, and then invokes \ro~$2K$ times by drawing additional $2K$ random numbers. Hence, the total cost of this phase is $(2K)\cdot O(\kappa^{3})+(5K)\cdot \RNG$, where $O(\kappa^{3})$ denotes the cost of modular exponentiation and \RNG~denotes the cost of drawing a random number. We omit the costs of XOR operations. 
 \item {\em Query phase:} \F draws $K$ random numbers to handle \A's \cemsig~queries, whose cost is  $K\cdot \RNG$. \F also draws $K'$ random numbers to handle \A's \ro~queries, whose cost is at most $K'\cdot \RNG$.
\end{itemize}
Therefore, the approximate total running time of \F is $t'=O(t)+(2K)\cdot O(\kappa^{3})+(6K+K')\cdot\RNG$.

\vspace{2mm}

 \noindent \underline{{\em Success Probability Analysis}}: 
 \F succeeds if all below events occur.

\begin{enumerate}[-]
\setlength{\itemsep}{0pt}
  \setlength{\parskip}{0pt}
  \setlength{\parsep}{0pt}
  \item \nab: \F does not abort during the query phase.

  \item \forge: \A wins the $K$-time \EUCMA~experiment for \somecs.

  \item \nabb: \F does not abort after \A's forgery.

\item   \suc: \F wins the $K$-time \EUCMA~experiment for \dl{\em-experiment}.

\item  $Pr[\suc] = Pr[\nab]\cdot Pr[\forge|\nab]\cdot Pr[\nabb|\nab \wedge \forge]$

  \end{enumerate}

 $\bullet$ {\em The probability that event \nab~occurs}: During the query phase, \F aborts if $(M_j||x_j)$ $\in\lh,~0\leq j \leq K-1$ holds, {\em before} \F inserts $(c_j||M_j)$ into \lh~(i.e., the simulation fails). This occurs if \A guesses the randomized output $c_j$ and then queries $(c_j||M_j)$ to \ro~{\em before} querying it to \somecssig. The probability that this occurs is $\frac{1}{2^{|q|}}$, which is negligible in terms of $\kappa$. Hence, $Pr[\nab]=(1-\frac{1}{2^{|q|}})\approx 1$.

 $\bullet$ {\em The probability that event \forge~occurs}: If \F does not abort, \A also does not abort since the simulated  view of \A is {\em indistinguishable} from the real view of \A (see the indistinguishability analysis). Therefore, $Pr[\forge|\nab]=\advsomecs$.

  $\bullet$ {\em The probability that event \nabb~occurs}: \F does not abort if the following conditions are satisfied:
\begin{enumerate} [i]
\item \A wins the \EUCMA~experiment for \somecs~on a message $M^{*}$ by querying it to \ro. The probability that \A wins without querying $M^{*}$ to \ro~is as difficult as a random guess.

\item After \F extracts $y'$ by solving modular linear equations, the probability that $Y' \not\equiv \alpha^{y'} \bmod p$ is negligible in terms $\kappa$, since $(Y=Y') \in \pk$ and $\somecsver(\pk,M^{*},\sigma^{*})=1$. Hence, $Pr[\nabb|\nab \wedge \forge]=\advsomecs$.
\end{enumerate}
  Omitting the terms that are negligible in terms of $\kappa$, the upper bound on {\em \EUCMA-advantage of \somecs}~is as follows:
\begin{eqnarray*}
\advcem & \le & \advdl,
\end{eqnarray*}

\vspace{2mm}
{\em \underline{Indistinguishability Argument}}: The real-view of \A is comprised of the public key $\pk=(Y,\alpha,p,q,\vv=( \langle \gamma_0,\beta_0 \rangle,\ldots, $ $\langle \gamma_{K-1},\beta_{K-1} \rangle,K)$, the answers of $\somecssig_{sk}(.)$ as $\lsigma=(s_j,c_j)$ for $j=0,\ldots,K-1$, and the answer of \ro~as $\lh=(h_0,\ldots,h_{K'-1})$ on corresponding $H_i, i\in\{0,1\}$, respectively. That is, $\lA_{\mathit{real}}=\langle \pk,\lsigma,\lh\rangle$, where all values are generated/computed by \somecs~algorithms as in the real system. All variables in \lA~are computed from the values $\{y,r_n,z_n,h_j,\alpha,p,q\}_{n=0,j=0}^{K-1,K'-1}$. Hence, the joint probability distribution of all other variables in \lA~are determined by the joint probability of these values. All these are random in $\mathbb{Z}_{q}^{*}$. Therefore, the joint probability distribution of \lA~is,
{\small
\begin{eqnarray*}
Pr[\Areal=\overrightarrow{a}] &= & Pr[\overline{y} = y | \overline{r}_0 = r_0 \wedge,\ldots,\overline{h}_{K'-1} = h_{K'-1}]\\
& = & \frac{1}{(q-1)^{1+2K+K}\cdot(p-1)^{2}}
\end{eqnarray*}
}
We denote the simulated view of \A is as \Asim, and it is equivalent to \Areal~except that in the simulation, values $(s_j,e_j,z_j,c_j)$ for $j=0,\ldots,K-1$ are randomly selected from $\mathbb{Z}_{q}^{*}$. Note that the joint probability distribution of these variables are identical to original signature and hash outputs (since hash function is modeled as RO). Hence, we write $Pr[\Areal=\overrightarrow{a}]=Pr[\Asim=\overrightarrow{a}]$.
\hfill$\square$

\begin{table*}[t!]
	\vspace{-2mm}
	\centering  \caption{Private/public key sizes, signature size and signature generation/verification costs of \somecs~and
		its counterparts} \label{tab:analytic}
	\vspace{-2mm}
	\begin{tabular}{|@{}c@{}||@{}c@{}|@{}c@{}|@{}c@{}||@{}c@{}|@{}c@{}|}
		
		\hline  \multicolumn{1}{|c||}{\multirow{2}{*} {\textbf{Scheme}}} & \multicolumn{3}{c||}{\bf{Signer}} & \multicolumn{2}{c|}{\bf{Verifier}} \\
		\cline{2-6}   & \specialcell[]{\textbf{Private Key} \\ \textbf{Size}} & \specialcell[]{\textbf{Signature} \\ \textbf{Size}} & \specialcell[]{\textbf{ Signature} \\ \textbf{ Generation}} & \specialcell[]{\textbf{Public Key} \\ \textbf{Size}} & \specialcell[]{\textbf{ Signature} \\ \textbf{ Verification}} \\ \hline \hline
		
		\multicolumn{6}{|c|}{\em \textbf{Full-time signatures}} \\ \hline
		
		SPHINCS~\cite{SPHINCS} &   $ n_S $   & \specialcell[]{ $ n_S(k_S({|t_S|} $ \\ $- x_S + 1) + 2^{x_S}) $ } & $ (2t_S - 1) \cdot H $  &   $  n_S  $ & \specialcell[]{$ (k_S((\log t_S) - x_S + 1)$ \\ $  + 2^{x_S} -1) \cdot H  $} \\ \hline

		ECDSA~\cite{ECDSA} & $|q|$ & $2|q|$ & $EMul$ &$|q|$ & $1.3\cdot EMul$ \\ \hline
		
		Ed25519~\cite{Ed25519} & $|q|$ & $2|q|$ & $EMul$ & $|q|$ & $1.3\cdot EMul$ \\ \hline 
		
		Kummer~\cite{Kummer} & $|q|$ & $2|q|$ & $EMul$ & $|q|$ & $1.3\cdot EMul$ \\ \hline
		
		SchnorrQ~\cite{SchnorrQ} & $|q|$ & $2|q|$ & $EMul$ & $|q|$ &  $1.3\cdot EMul$\\ \hline  \hline 
		
		\multicolumn{6}{|c|}{\em \textbf{$K$-time signatures}} \\ \hline
		
		HORS~\cite{HORS_BetterthanBiBa02}  & $\kappa$ & $\kappa \cdot u$ & $ (u + 1) \cdot H$ & $t\cdot|H| \cdot K$ & $(u+1)\cdot H$ \\ \hline
		
		HORSE~\cite{OTS_r_time_HORSED_2006}  & $(\kappa \cdot t \cdot log_{2}(K))$ & $\kappa \cdot u$ &~ $(u \cdot log_{2}(K)+1)\cdot H$ &$t\cdot|H|$ & $(u+1)\cdot H$ \\ \hline
		
		
		XMSS~\cite{XMSS} & $\kappa$ & \specialcell[]{$(l + log_{2}(K)) $ \\ $ \cdot |H|$} & \specialcell[]{$(((log_{2}(K)+2)\cdot (log_{2}(K)$ \\ $ + l \cdot (w+2)))/2 + 4 \cdot log_{2}(K)) \cdot H$} & \specialcell[]{$(2(log_{2}(K) + log_2(l))$ \\ $+1) \cdot |H|$} & \specialcell[]{$(log_{2}(K)+ l $\\ $\cdot (w + 1)) \cdot H $} \\ \hline 
		
		Zaverucha et al.~\cite{Zaverucha} & $\kappa$ &  $\kappa + |q|$& $(m/2) \cdot (Add_q + H)$ & $m \cdot |q| \cdot K$& $1.3\cdot EMul$ \\ \hline \hline
		
		\somecs & $|q|$ & $|q|$ & $2\cdot H + Mul_q + Sub_q$ & $(2K+1)\cdot |q|$& $1.3\cdot EMul$ \\ \hline 
		
	\end{tabular}
	\flushleft{\scriptsize{
		
		$K$ denotes the number of signatures that can be generated using a single key pair for $K$-time signature schemes.
		
		$Emul$ and $Eadd$ denote the costs of EC scalar multiplication over modulus $p'$, and EC addition over modulus $p'$, respectively. ECDSA~\cite{ECDSA}, Ed25519~\cite{Ed25519}, Kummer~\cite{Kummer}, and SchnorrQ~\cite{SchnorrQ} only differ from each other in terms of the underlying curve. The operations that are required are the same for these schemes. $H$ and $Mul_{q}$ denote a cryptographic hash and a modular multiplication over modulus $q$, respectively.  We omit the constant number of negligible operations if there is an expensive operation (e.g., integer additions are omitted if there is an $Emul$). We use double-point scalar multiplication for verifications of ECC based schemes ($1.3 \cdot Emul$ instead of $2 \cdot Emul$~\cite{ECCGuide}). $t_S$, $k_S$ and $x_S$ are SPHINCS~\cite{SPHINCS} parameters where $t_S$ is the number of secret key elements, $k_S$ is the number of revealed secret key elements and $x_S$ is a small integer.	SPHINCS~\cite{SPHINCS} parameter $n_S$ denotes the bit length of hashes. Zaverucha et al.~\cite{Zaverucha} parameter $m$ should be selected such that $m\choose m/2$ $ \geq 2^{2\kappa}$. Integers $t$ and $u$ denote the parameters used in HORS~\cite{HORS_BetterthanBiBa02} and HORSE~\cite{OTS_r_time_HORSED_2006}. $w$ is the Winternitz parameter and $l$ is the tree parameter in XMSS~\cite{XMSS}.
		
		
		
		
		\textbf{Remark:} For HORS~\cite{HORS_BetterthanBiBa02} and Zaverucha et al.~\cite{Zaverucha}, similarly to \somecs, we deterministically generate the necessary private key components from a seed (i.e., using a keyed hash) to have a small constant private key that can be deployed to low-end devices.

	}
	}
\vspace{-5mm}
\end{table*}

\section{Performance Analysis and Comparison}
\label{sec:PerformanceAnalysis}

In this section, we first present the analytical analysis of \somecs~and its counterparts. Then, we present the results of our experiments on a commodity laptop and an 8-bit AVR embedded processor. Our evaluation metrics include key sizes, signature size, and computation costs. On the 8-bit microprocessor, we focus on the signer cost (i.e., private key, signature size, signature generation) since our system model includes resource-constrained devices as signers. For our counterparts, we consider state-of-the-art $K$-time signature schemes as well as some traditional (full-time) signatures. 

\vspace{1mm}
\noindent \textbf{Remark:} Our envisioned applications require high signer efficiency to be practical on resource-constrained devices. Hence, {\em optimizing the online signer efficiency is the essential performance objective for \somecs}. Recall that we assume verifiers are resourceful entities, which is a reasonable assumption for our envisioned applications (see Section \ref{sec:Introduction}). Also note that in \somecs~system model, private/public keys are generated before the system deployment (see Section \ref{subsec:SecModel}). Hence, the key generation cost (i.e., {\em offline cost}) is not a critical performance metric for \somecs.

\subsection{Analytical Performance Analysis}

Here, we describe the analytical costs of our scheme, where the online costs are summarized in Table~\ref{tab:analytic}.

\noindent \textbf{Key Generation: }Key generation of \somecs~requires $K$ EC scalar multiplications that is higher than its full-time counterparts. For instance, for EC-based signature schemes (e.g., ECDSA~\cite{ECDSA}, Ed25519~\cite{Ed25519}, Kummer~\cite{Kummer}, and SchnorrQ~\cite{SchnorrQ}) keys are generated with only one EC scalar multiplication. However, it is comparable to its $K$-time counterparts as their key generation also depends on $K$. Note that in our system model, key distribution is performed before the deployment. Thus, we believe that this does not pose a limitation for our considered use-cases.

\noindent \textbf{Signer Overhead: }In \somecs,~signer stores a small private key that is the same size as its full-time elliptic curve counterparts. The private key of some $K$-time signatures can be deterministically derived from a $\kappa$-bit seed, which is 2$\times$ smaller than that of \somecs. However, this makes a small difference in practice (i.e., 16 Bytes vs 32 Bytes). Signature generation of \somecs~only requires 2 hash function calls, a single multiplication, and subtraction under$\mod q$. This introduces a significantly smaller overhead compared to its alternatives. The counterparts of \somecs~either require expensive operations (i.e., EC scalar multiplication) or a very large number of hash function calls for signature generation. Only HORS~\cite{HORS_BetterthanBiBa02} and Zaverucha et al.~\cite{Zaverucha} have comparable signature generation speed. However, when we generate the private key components from a seed, these hash function calls dominate the signature generation cost for these schemes due to their large private key size. 

\noindent \textbf{Signature Transmission: }Signature size of \somecs~is the smallest compared to its counterparts. Note that since the signature component $c_j$ contains the information to recover the first $|q|$ Bytes of the message, we do not consider its transmission as an overhead. \somecs~only requires additional $|q|$ Bytes to be transmitted. Since the transmission of signatures introduces an overhead to the energy consumption of signer (and verifier), we believe it is essential to minimize its size.

\noindent \textbf{Verifier Overhead: }In \somecs~the public key is linear with the messages to be signed with a single key pair. Therefore, it increases as $K$ increases (similar to its $K$-time counterparts except HORSE~\cite{OTS_r_time_HORSED_2006}). Considering that the verifier device is a resourceful device (e.g., server, command center) in \somecs~applications, we believe this is tolerable. The signature verification of \somecs~requires an EC double scalar multiplication (can be accelerated with Shamir's trick~\cite{ECCGuide}). 

\noindent \textbf{Parameters: } {\em We selected parameters to provide $\kappa = 128$-bit security for both \somecs~and its counterparts.} For elliptic-curve based schemes (including \somecs), we selected $|q| = 256$-bit. For Zaverucha et al., we selected $m=260$, for HORS and HORSE, we selected $t = 1024$ and $u = 24$ to provide the desired security level. For XMSS and SPHINCS, we used the parameters suggested in the base papers. We refer the interested readers to the base papers of these schemes for the detailed explanation of their parameter choice.

\begin{table*}[t!]
	\centering
	\vspace{-2mm}
	\caption{Experimental performance comparison of \somecs~and its counterparts on a commodity hardware} \label{tab:Laptop}
	\vspace{-2mm}
	\begin{threeparttable}
		\begin{tabular}{| c || c | c | c | c | c |  c | c | }
			\hline
			\textbf{Scheme} & $K$ & \specialcell[]{\textbf{Signature Generation}\\  \textbf{ Time (}CPU cycle\textbf{)}} & \specialcell[]{\textbf{Private Key}\textsuperscript{$ \mathparagraph$} \\ \textbf{(Byte)}} & \specialcell[]{\textbf{Signature }\\  \textbf{Size (Byte)}} & \specialcell[]{\textbf{Signature Verification}\\  \textbf{ Time (}CPU cycle\textbf{)}} & \textbf{Public Key}\textsuperscript{$\ddagger$}  & \specialcell[]{\textbf{End-to-End} \\ \textbf{Delay (}CPU cycle\textbf{)}} \\ \hline \hline 
			
			
			\multicolumn{8}{|c|}{\em \textbf{Full-time signatures}} \\ \hline
			
			SPHINCS~\cite{SPHINCS} & $2^{\kappa}$& 37 466 005 & 1088 & 41000 & 1 051 562 & 1056 & 38 517 567 \\ \hline
			
			
			ECDSA~\cite{ECDSA} & $2^{\kappa}$ & 1 510 320 & 32 & 64 &  1 932 650 & 32 & 3 442 970 \\ \hline
			
			Ed25519~\cite{Ed25519} & $2^{\kappa}$ & 146 620 & 32 & 64 & 286 750 & 32 & 433 370 \\ \hline
			
			Kummer~\cite{Kummer} & $2^{\kappa}$& 58 450 & 32 & 64 & 98 560 & 32 & 157 010 \\ \hline 
			
			SchnorrQ~\cite{SchnorrQ} & $2^{\kappa}$ & 30 481 & 32 & 64 & 54 241 & 32 & 84 722 \\ \hline \hline
			
			\multicolumn{8}{|c|}{\em \textbf{$K$-time signatures}} \\ \hline
			
			\multirow{2}{*}{HORS~\cite{HORS_BetterthanBiBa02}} & 1 & \multirow{2}{*}{16 823} & \multirow{2}{*}{16} & \multirow{2}{*}{384} & \multirow{2}{*}{8 975} & 32 KB & \multirow{2}{*}{25 798} \\ 
			
			
			& $2^{17}$& & & & & 4 GB & \\ \hline

			\multirow{2}{*}{HORSE~\cite{OTS_r_time_HORSED_2006}} & 1 & 16 823 & 16384 & \multirow{2}{*}{384} &  \multirow{2}{*}{8 975} & \multirow{2}{*}{32 KB} & 25 798 \\ 
			
			
			& $2^{17}$& 280 247 & 278 528 & & & & 287 503 \\ \hline 
			
			\multirow{2}{*}{XMSS~\cite{XMSS}} & 1 & 137 856 & \multirow{2}{*}{16} & 2080 & 115 239 & 416 & 253 095 \\ 
			
			
			& $2^{17}$& 1 367 431 & & 2592 & 120 983 & 1504 & 1 488 414 \\ \hline
			
			\multirow{2}{*}{Zaverucha et al.~\cite{Zaverucha}} & 1 & \multirow{2}{*}{89 180} &  \multirow{2}{*}{16} & \multirow{2}{*}{48} & \multirow{2}{*}{52 872} & 4160 & \multirow{2}{*}{142 052} \\ 
			
			
			& $2^{17}$& & & & & 520 MB & \\ \hline \hline 
			
			
			\multirow{2}{*}{\somecs} & 1 & \multirow{2}{*}{\textbf{2 425}} & \multirow{2}{*}{\textbf{32}} & \multirow{2}{*}{\textbf{32}} & \multirow{2}{*}{52 872} & 96 & \multirow{2}{*}{\textbf{55 297}} \\ 
			
			
			& $2^{17}$&  &  & & & 8 MB & \\ \hline
			
		\end{tabular}
		\begin{tablenotes}[flushleft]\scriptsize{  
				\item $\ddagger$ The sizes are in terms of \textbf{Bytes}, if otherwise not specified.
				
				\item $ \mathparagraph $ System wide parameters \params~(e.g., p,q,$\alpha$) for each scheme are included in their corresponding codes, and private key size denote to specific private key size.
				
				\item The cost of hash-based schemes are estimated based on the cost of a single hash operation.
				
			}
		\end{tablenotes}
	\end{threeparttable}
	\vspace{-3mm}
\end{table*}

\subsection{Performance Evaluation on Commodity Laptop}

We implemented \somecs~on a laptop and compared its cost to its state-of-the-art counterparts. 

\noindent \textbf{Hardware Configurations and Software Libraries: } As our commodity hardware, we used a laptop equipped with an Intel i7 Skylake 2.6 GHz CPU with 12 GB RAM.

We implemented \somecs~on FourQ curve~\cite{FourQ} to offer fast verification. We used the open-source implementation of this curve which can be found at\footnote{\url{https://github.com/Microsoft/FourQlib}}. We used our hash function as blake due to its high efficiency and high security~\cite{blakeHash}. Specifically, we used blake2s due to its optimization on low-end devices. We open-source our implementations at

\begin{center}
	\fbox{\url{www.github.com/ozgurozmen/SEMECS}}
\end{center}

We ran the open-sourced implementations of our counterparts on our hardware setting, if possible. For the hash-based constructions, we conservatively simulated their costs with blake2s hash function, to be fair with them. 

\noindent \textbf{Experimental Results: } Table~\ref{tab:Laptop} shows the benchmarks and specific key/signature sizes for \somecs~and its counterparts. We observe that \somecs~is 7$\times$ faster than its closest counterpart (HORS~\cite{HORS_BetterthanBiBa02}) in terms of signature generation. Specifically, {\em it takes only $1.23$ microseconds to generate a signature with \somecs}. Moreover, it has a compact private key of 32 Bytes and the smallest signature (32 Bytes) among its counterparts. Signature verification of \somecs~is also fast since we used the optimized FourQ~\cite{FourQ} curve to implement our scheme. Therefore, only HORS~\cite{HORS_BetterthanBiBa02} and HORSE~\cite{OTS_r_time_HORSED_2006} offer faster verification. The main limitation of \somecs~is its public key size. Specifically, when $K = 2^{17}$, which { \em allows signing a message in every 20 minutes for 5 years without a key replacement}, the public key size is 8 MB. However, this is much smaller than some of the most efficient $K$-time counterparts such as HORS, and Zaverucha et al., that have 4 GB and 520 MB public key, respectively. We also implemented the key generation of \somecs~on this experimental setting and observed that generating the key for $K= 2^{17}$ takes $1.75$ seconds. 

\subsection{Performance/Energy Evaluation on 8-bit Microprocessor}

We fully implemented the signature generation of \somecs~on an 8-bit microprocessor to assess its energy and time efficiency on low-end embedded devices.

\noindent \textbf{Hardware Configurations and Software Libraries: } We used 8-bit AVR ATmega 2560 microprocessor to measure the signer efficiency of our scheme compared to its counterparts. We selected this low-end device due to its low energy consumption and extensive use in practice, especially in IoT applications and medical devices~\cite{IMDProcessor, ATmega2560Medical, ATmega2560Medical2}. It is equipped with 256 KB flash memory, 8 KB SRAM, 4 KB EEPROM, and its maximum clock frequency is 16 MHz.

We implemented \somecs~using Rhys Weatherley's crypto library~\cite{weatherley} that offers high-speed operations for low-end devices. Specifically, we used its blake2s implementation and modified its reduction algorithm to compute$\mod q$ (where $q$ is FourQ parameter) using Barrett reduction. We also open-source our 8-bit implementations at the link given above to facilitate the test and broad adoption of \somecs. 

We used the results of our counterparts that were given in 8-bit AVR microprocessors, if possible. For instance, we used the results provided in~\cite{Ed255198bit} for Ed25519,~\cite{Kummer8bit} for $\mu$Kummer and~\cite{FourQ8bit} for SchnorrQ. We ran the ECDSA implementation of microECC~\cite{microECC} on our hardware. Similar to the laptop implementation, we measured the cost of a single hash (blake2s) call on our microprocessor and conservatively estimated the hash-based schemes' cost. 

\noindent \textbf{Experimental Results: }As summarized in Table~\ref{tab:AVR}, our analysis confirmed that \somecs~is highly efficient at the signer's side. Signature generation of \somecs~is performed with less than 200 thousand cycles, which is 6$\times$ faster than HORS~\cite{HORS_BetterthanBiBa02} and 19$\times$ faster than SchnorrQ (its fastest counterparts). In addition to this, \somecs~requires a small private key and a signature size that is the smallest among its counterparts. This makes \somecs~very desirable for applications that include resource-limited signers. 

\begin{figure*}[t]
	\centering
	\begin{subfigure}[t]{.5\textwidth}	
		\includegraphics[width=\linewidth]{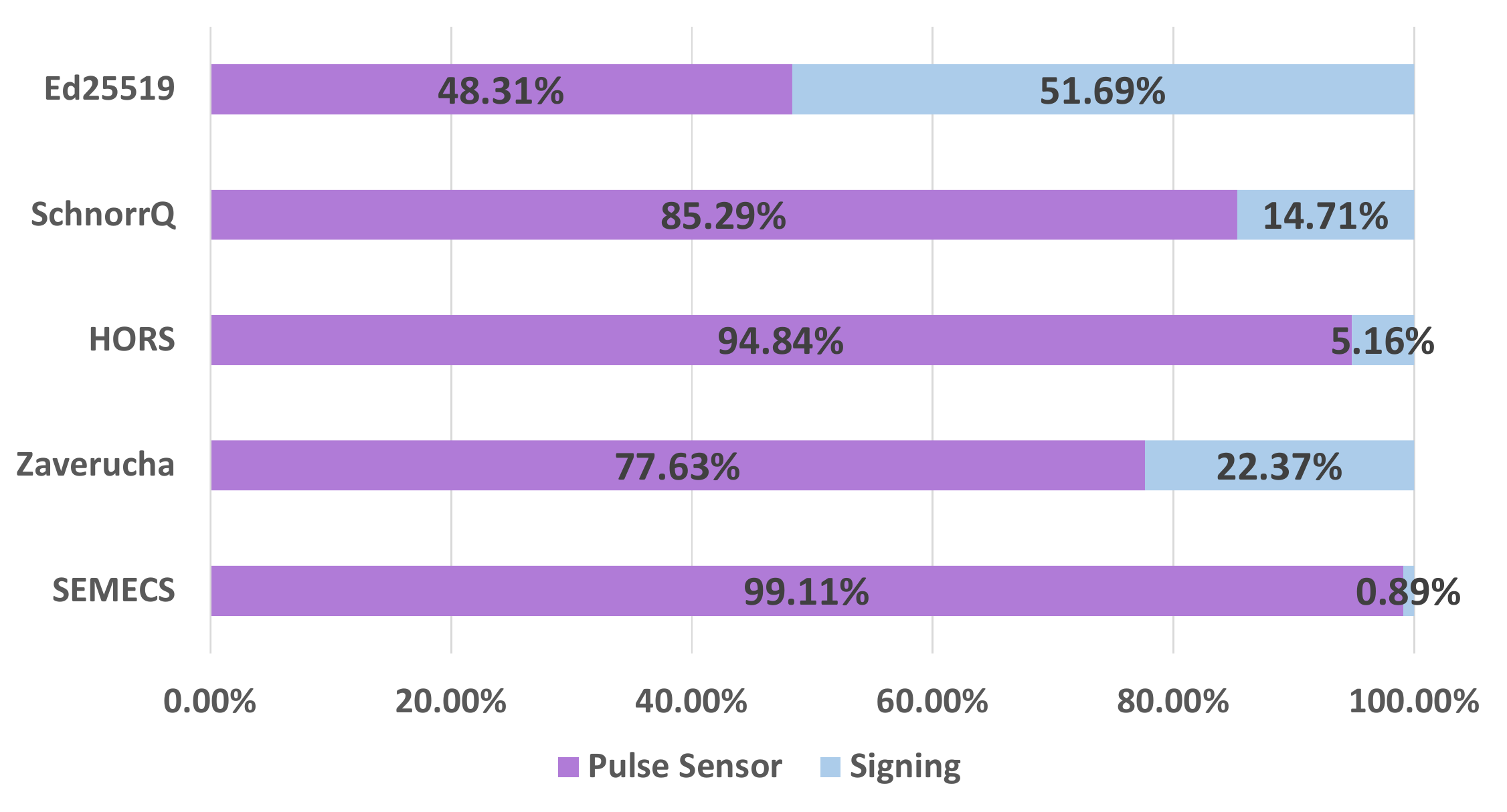}
		\caption{Energy of Signature Generation vs Pulse Sensor}
		\label{fig:areaPulse}
		\vspace{-1mm}
	\end{subfigure}%
	\begin{subfigure}[t]{.5\textwidth}		
		\includegraphics[width=\linewidth]{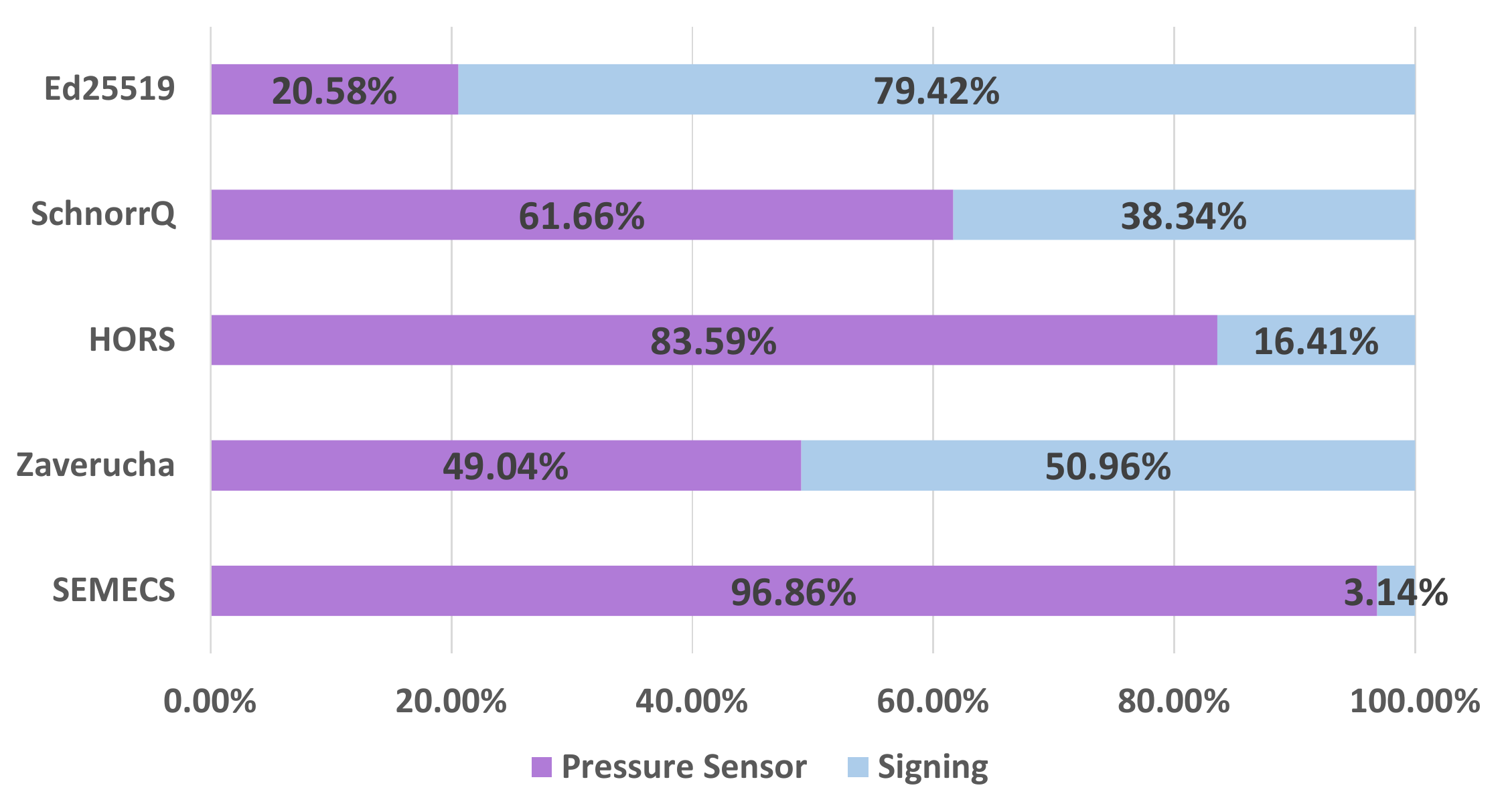}
		\caption{Energy of Signature Generation vs Pressure Sensor}
		\label{fig:areaPressure}
		\vspace{-1mm}
	\end{subfigure}
	\caption{Energy consumption of signature generation vs IoT sensors} \label{fig:areaBattery}
	\vspace{-4mm}
\end{figure*}

\noindent \textbf{Energy Consumption Analysis: }We analyzed the energy consumption of \somecs~and its counterparts on our experimental setting and compared with the energy consumption of two common IoT sensors (a pulse and a pressure sensor). We first derived a generic energy consumption estimation (as in~\cite{MICAZenergy} that offers an estimation for MICAz) for 8-bit AVR ATmega 2560 based on our \somecs~implementation and used it to estimate the energy consumption of our counterparts (similarly with~\cite{FourQ8bit} that uses~\cite{MICAZenergy}). We also calculated how much energy is required to operate IoT sensors. We took into consideration (i) energy drawn by the sensor (ii) energy drawn by the microprocessor to read data from the sensor and (iii) energy drawn by the microprocessor during the idle time.

We powered our microprocessor with a 2200 mAh power pack. This allowed us to use an ammeter/power meter connected between the battery and the microprocessor. We measured 5V of voltage and 20mA of current on load, which is verified by the datasheet of the processor\footnote{\url{http://www.atmel.com/Images/Atmel-2549-8-bit-AVR-Microcontroller-ATmega640-1280-1281-2560-2561\_datasheet.pdf}}. Then, we used the formula $E = V \cdot I \cdot t$ to calculate the energy consumption in Joules (as in~\cite{Precomputation:LowCostSig:Ateniese:NDSS2013}). We also considered the potential deployment of nRF24L01 Single Chip 2.4 GHz Transceiver to 8-bit ATmega for signature transmission. Based on its datasheet, we also estimated the energy consumption of signature  transmission with this low-power transceiver. Specifically, nRF24L01 operates at 3.3 V, 11.3 mA and support a transmission rate of 2Mbps. {\em Our results showed that 8-bit AVR ATmega 2560 consumes roughly $6.25nJ$ per cycle of computation and $18.65nJ$ per bit of transmission.}

We also calculated how much energy is necessary to operate IoT sensors. Specifically, we used a pulse sensor\footnote{\url{https://pulsesensor.com/}} (that could serve as an example of a medical sensor) and a BMP183 pressure sensor\footnote{\url{https://cdn-shop.adafruit.com/datasheets/1900_BMP183.pdf}} (that could be an example of a daily IoT application). In our energy calculations, we considered a sampling frequency of 10 seconds for the pulse sensor and 10 minutes for the pressure sensor, due to the difference/urgency in their usage. Figure~\ref{fig:areaBattery} shows how many percentage of the battery is spent on the IoT sensor, compared with that of the cryptographic operations (i.e., signing) of different schemes. One can observe that for pulse sensor (see Figure~\ref{fig:areaPulse}), HORS and SchnorrQ require the $5.16\%$ and $14.71\%$, whereas, with \somecs, this is decreased to a negligible level ($0.89\%$). For the pressure sensor, the energy consumption of \somecs~is only $3.14\%$ where the closest counterpart is $16.41\%$. This shows that preferring \somecs~as the authentication mechanism in 8-bit AVR microprocessors significantly reduces the impact of cryptographic operations on battery life.

Based on this analysis, we noticed that \somecs~outperforms its counterparts for both computation energy and communication energy at the signer's side. We believe that this is essential in practice to extend the battery lives of critical embedded devices such as implantable medical devices.

\section{Conclusion}
\label{sec:Conclusion}

In this paper, we proposed a new signature scheme called \somecs, which achieves several desirable properties that are critical for resource-constrained devices. Specifically, \somecs~only requires two hash, a modular multiplication, and a modular subtraction to compute a signature. Moreover, it has a constant-small private key and signature, that is optimal for an EC-based signature scheme. Our experiments on both laptop and 8-bit AVR confirmed the energy and computational efficiency of \somecs. Therefore, we believe \somecs~is an ideal alternative for providing authentication and integrity services for resource-constrained devices. 

\section*{Acknowledgments}
\addcontentsline{toc}{section}{Acknowledgments}

This work is supported by the NSF CAREER Award CNS-1652389. We would like to thank Rouzbeh Behnia for his valuable comments.

\bibliographystyle{IEEEtran}
\bibliography{crypto-etc}

\begin{IEEEbiography}[{\includegraphics[width=1in, height=1.25in, clip,keepaspectratio]{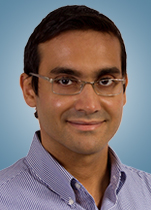}}]{Attila Altay Yavuz}
	(M `11) is an Assistant Professor in the Department of Computer Science and Engineering, University of South Florida (August 2018). He was an Assistant Professor in the School of Electrical Engineering and Computer Science, Oregon State University (2014-2018). He was a member of the security and privacy research group at the Robert Bosch Research and Technology Center North America (2011-2014). He received his PhD degree in Computer Science from North Carolina State University in August 2011. He received his MS degree in Computer Science from Bogazici University (2006) in Istanbul, Turkey. He is broadly interested in design, analysis and application of cryptographic tools and protocols to enhance the security of computer networks and systems. Attila Altay Yavuz is a recipient of NSF CAREER Award (2017). His research on privacy enhancing technologies (searchable encryption) and intra-vehicular network security are in the process of technology transfer with potential world-wide deployments. He has authored more than 40 research articles in top conferences and journals along with several patents. He is a member of IEEE and ACM.
\end{IEEEbiography}
\vskip 0pt plus -1fil
\begin{IEEEbiography}[{\includegraphics[width=1in, height=1.25in, clip,keepaspectratio]{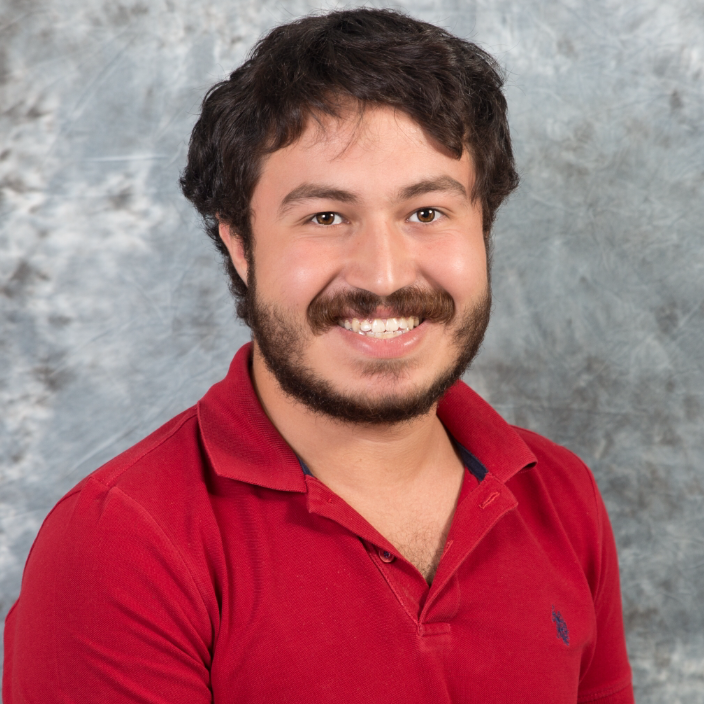}}]{Muslum Ozgur Ozmen}
	received the bachelor`s degree in electrical and electronics engineering from the Bilkent University, Turkey and the M.S. degree in computer science from Oregon State University. He is currently pursuing a PhD degree in computer science with the Department of Computer Science and Engineering, University of South Florida. His research interests include lightweight cryptography for IoT systems (drones and medical devices), digital signatures, privacy enhancing technologies (dynamic symmetric and public key based searchable encryption) and post-quantum cryptography.
\end{IEEEbiography}

\end{document}